\documentclass[useAMS,usenatbib]{mn2e}

\def\Msol{\hbox{M$_\odot$}}

\def\kms{\hbox{km$\,$s$^{-1}$}}
\def\cmt{\hbox{cm$^{-3}$}}

\def\two{\,{\sc ii}}
\def\three{\,{\sc iii}}

\newcommand\fsec{\hbox{$.\!\!^{\rm s}$}}

\usepackage{amsmath}
\usepackage{amssymb}
\usepackage{mathrsfs}
\usepackage{upgreek}
\usepackage{graphicx}
\usepackage[usenames,dvipsnames]{color,colortbl}
\usepackage{soul} 
\usepackage{overpic}
\usepackage{threeparttable}
\usepackage{tikz} 
\newcommand*\circled[1]{\tikz[baseline=(char.base)]{
            \node[shape=circle,draw=white,inner sep=1pt] (char) {#1};}}

\title[IFU observations of gas pillars in NGC 3603]{Optical IFU observations of gas pillars surrounding the super star cluster NGC 3603}
\author[M.\ S.\ Westmoquette et al.] {M.\ S.\ Westmoquette$^1$\thanks{E-mail: westmoquette@gmail.com}, J.\ E.\ Dale$^{2,3}$, B.\ Ercolano$^{2,3}$, L.\ J.\ Smith$^4$ \\
$^1$European Southern Observatory, Karl-Schwarzschild-Str. 2, 85748 Garching bei M\"{u}nchen, Germany \\
$^2$Excellence Cluster `Universe', Boltzmannstra\ss e 2, 85748 Garching, Germany \\
$^3$Universit\"{a}ts-Sternwarte M\"{u}nchen, Scheinerstra\ss e 1, 81679 M\"{u}nchen, Germany \\
$^4$Space Telescope Science Institute and European Space Agency, 3700 San Martin Drive, Baltimore, MD 21218, USA\\
}
\date{}
\pagerange{\pageref{firstpage}--\pageref{lastpage}}
\pubyear{2013}
\begin{document}

\defcitealias{westm10b}{Paper~I}

\maketitle
\label{firstpage}
\begin{abstract}
We present optical integral field unit (IFU) observations of two gas pillars surrounding the Galactic young massive star cluster NGC 3603. The high S/N and spectral resolution of these data have allowed us to accurately quantify the H$\alpha$, [N\two] and [S\two] emission line shapes, and we find a mixture of broad (FWHM$\sim$70--100~\kms) and narrow ($<$50~\kms) components. The broad components are found close to the edges of both pillars, suggesting that they originate in turbulent mixing layers (TMLs) driven by the effect of the star cluster wind. Both pillars exhibit surprisingly high ionized gas densities of $>$10\,000~\cmt. In one pillar we found that these high densities are only found in the narrow component, implying it must originate from deeper within the pillar than the broad component. From this, together with our kinematical data, we conclude that the narrow component traces a photoevaporation flow, and that the TML forms at the interface with the hot wind. On the pillar surfaces we find a consistent offset in radial velocity between the narrow (brighter) components of H$\alpha$ and [N\two] of $\sim$5--8~\kms, for which we were unable to find a satisfactory explanation. We urge the theoretical community to simulate mechanical and radiative cloud interactions in more detail to address the many unanswered questions raised by this study.
\end{abstract}

\begin{keywords} ISM: H\two\ regions -- ISM: individual: NGC 3603 -- ISM: kinematics and dynamics.
\end{keywords}

\section{Introduction}\label{intro}

Massive stars dominate the energetics of the ISM due to their powerful radiative and mechanical feedback effects. In the simple case of a single star or cluster embedded in a uniform density and pressure medium, photoionization, stellar winds and subsequent supernova(e) ejecta act to inflate a spherical bubble of hot shocked gas, sweeping the ISM material into a thin dense shell \citep{weaver77}. This model breaks down, however, when the surrounding medium is not uniform, but fragmented into dense clumps -- as is observed in reality.

Gas clumps can have dramatic effects on the evolution of the wind, both in the case of a single star and a star cluster-driven galactic wind. Acting as obstacles, they deflect the flow, inducing shocks and therefore immediately enhancing radiative energy losses. Furthermore the clumps disintegrate through photoevaporation and ablation \citep{pittard07, gritschneder09, gritschneder10}, and the resultant entrainment and/or loading of matter into the flow causes additional downstream radiative energy losses \citep{fierlinger12, rogers13}. A full knowledge of the location and degree of the energy exchanges and losses is therefore essential due to their critical effect on the evolution of the flow.

The selective erosion or ablation of an inhomogeneous density field by radiation or stellar winds naturally results in elephants trunks or pillars pointing towards the source(s) of feedback. Denser-than-average gas resists destruction, at least for a time, and shields material downstream, forming elongated structures. The density inhomogeneities required can be in the form of well-defined isolated clumps \citep{williams01, miao06}, perturbations in ionization fronts \citep{tremblin12}, filamentary structures or accretion flows \citep{dale12}, fractal density distributions \citep{walch12} or turbulent density fields \citep{gritschneder10, tremblin12b}.

Pillars are a common feature of regions of the ISM experiencing feedback from OB-stars \citep[e.g.][]{sugitani02, rathborne04, schneider12}. The variety of formation mechanisms for pillars makes it difficult to infer how any given such structure came about. However, pillars lie on the interface between hot photoionized or wind-blown gas generated by massive stars, and the cold undisturbed material of the host molecular clouds. Their study is essential to help address outstanding questions such as how efficient are massive stars in triggering star formation \citep[e.g.][]{dale12}, and what is the most important form of stellar feedback in different environments or ISM conditions.

In \citet[][hereafter \citetalias{westm10b}]{westm10b} we presented optical/near-IR integral field unit (IFU) observations of a gas pillar (approximating a clump) in the Galactic H\two\ region NGC~6357 ionized by the young open star cluster Pismis 24 \citep[containing 9 O stars;][]{fang12}. We identified narrow ($\sim$20~\kms) and broad (50--150~\kms) ionized gas emission line components dynamically associated with the pillar surface, and, based on previous work \citep{falgarone90, hartquist92, westm07a, binette09}, concluded that the broad component arises in a turbulent mixing layer \citep[TML;][]{slavin93} on the pillar surface set up by the impact of the stellar winds. This broad component thus provides us with a direct probe of the elusive wind-ISM interaction. These broad ionized emission line components are seen almost ubiquitously in nearby starburst environments \citep{marlowe95, mendez97, james09, westm07a, westm09a, westm11}, and it was the tentative results from these studies that formed the motivation for the study in \citetalias{westm10b}. 


However, to allow us to relate the results of \citetalias{westm10b} more directly to these types of more violent starbursting environments, we must also study wind-ISM interactions in higher energy cases. Here we present observations of two pillars surrounding the much more massive (and luminous) star cluster in the Galactic giant H\two\ region NGC~3603. NGC~3603, located at a distance of $7\pm1$~kpc \citep[1~arcmin = 2~pc;][]{moffat83, melena08}, is powered by the young, bright, compact stellar cluster (HD 97950). The three WNL, six O3, and numerous other late O-type stars contribute to a bolometric luminosity $\sim$100 times that of the Orion and Pismis 24 clusters and 0.1 times that of NGC~2070 (the central cluster in the 30 Dor complex in the Large Magellanic Cloud). Its age of a few~Myr \citep{sung04, beccari10} and total mass of $>$$10^4$~\Msol\ \citep[][and references therein]{harayama08} classify it as a super star cluster, and therefore much more analogous to a starburst environment. The HD 97950 star cluster lies in a wind-blown cavity north of a large molecular cloud \citep{clayton86, melnick89, nurnberger02}. The gaseous surroundings of the cluster show a complex and variable velocity, density and extinction structure \citep{drissen95, nurnberger03, pang11}, and include the two prominent pillars studied here. These pillars are approximately the same distance from the cluster as the one in NGC~6357 studied in \citetalias{westm10b} ($\sim$1--1.4~pc\footnote{There was a mistake in \citetalias{westm10b}: At a distance of 2.56~kpc for NGC~6357 \citep{massey01}, 10~arcsec = 0.12~pc and the separation between the star cluster and the pillar tip is therefore 0.7~pc, or at a distance of 1.7~kpc \citep{fang12} the cluster-pillar separation is 1.2~pc. Both values are smaller than the 3.5~pc quoted in \citetalias{westm10b}.}).


%


\section{Observations} \label{sect:obs}

\subsection{HST imaging}

\begin{figure*}
\includegraphics[width=\textwidth]{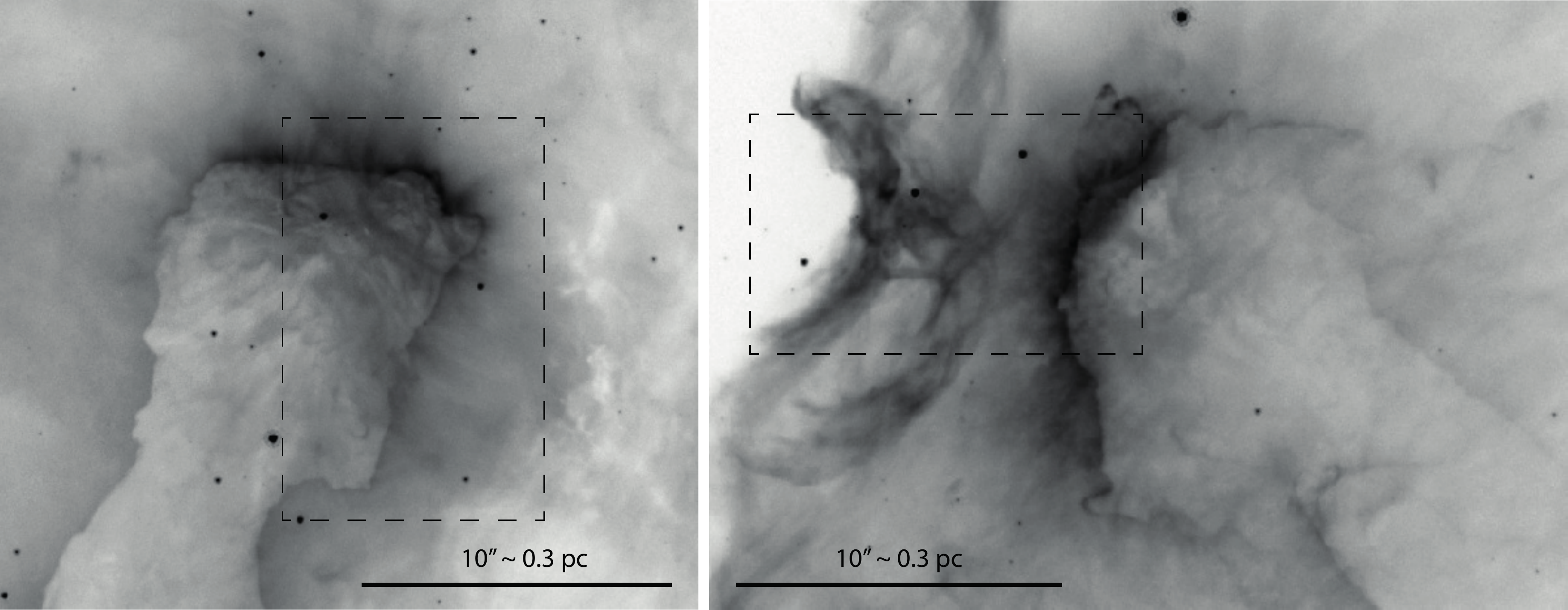}
\caption{Zoom-ins of the \textit{HST}/WFC3-UVIS F656N image showing the two pillar tips. The dashed rectangles indicate the location of the two ARGUS IFU pointings. The image grey-scales are inverted and square-root scaled to show the full dynamic range in each region.}
\label{fig:zooms}
\end{figure*}

We obtained \textit{HST}/WFC3-UVIS F656N (H$\alpha$+continuum) imaging of NGC~3603 from the \textit{HST} archive (PID: 11360, PI O'Connell). This is shown in Fig.~\ref{fig:finder}. Zoom-ins of the two pillars studied here are shown in Fig.~\ref{fig:zooms} with an enhanced colour scaling to show the full dynamic range of the image. The position 2 pillar (left) looks remarkably like the pillars in M16 \citep{hester96}, and, like in M16, the linear striations oriented normal to the pillar surface offer clear morphological evidence for photoevaporative flows. These linear striations are not evident on the position 1 pillar.

\subsection{IFU data}

\begin{figure*}
\includegraphics[width=\textwidth]{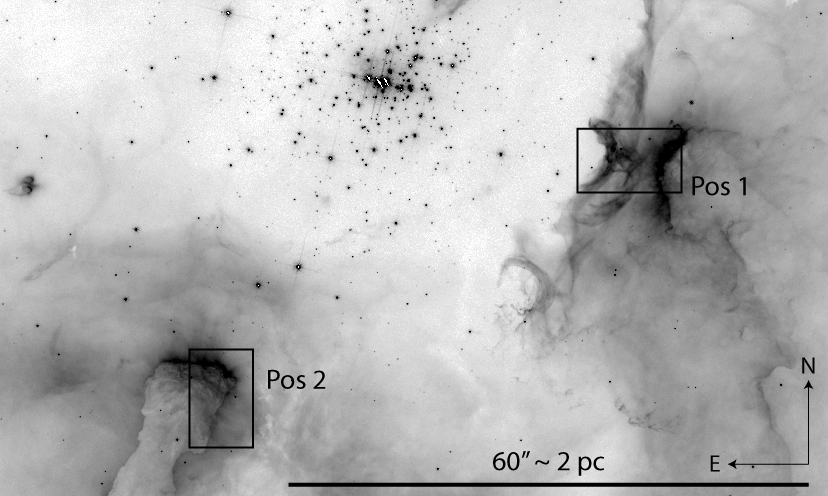}
\caption{\textit{HST}/WFC3-UVIS F656N (H$\alpha$+continuum) image of NGC~3603 showing the two ARGUS IFU pointings. The grey-scale colour map is inverted and square-root scaled.}
\label{fig:finder}
\end{figure*}

Observations of two pillars in NGC 3603 were obtained with the FLAMES instrument on the VLT coupled to the GIRAFFE spectrograph (PI: Westmoquette, Prop IDs: 087.C-0106, 089.C-0016). We used the ARGUS IFU array in its 0.52 arcsec/spaxel spatial sampling mode, giving a field-of-view (FoV) of $11.5\times 7.3$ arcsec sampled by 333 fibres. The observations were centred on the H$\alpha$-bright tips of two pillars, with coordinates Pos1 = $11^{\rm h}\,15^{\rm m}\,03\fsec5$, $-61^{\circ}\,15'\,47\farcs4$ and Pos2 = $11^{\rm h}\,15^{\rm m}\,09\fsec9$, $-61^{\circ}\,16'\,15\farcs5$ (J2000). Fig.~\ref{fig:finder} shows the position of the IFU positions on an \textit{HST}/WFC3-UVIS F656N (H$\alpha$) image of NGC 3603 (PID: 11360, PI: O'Connell). We observed both positions with the LR06 (L682) grating, giving a wavelength coverage of 6450--7200~\AA. A total  time of 90~mins for each position was split into two equal-length exposures, which were spatially dithered with an offset of $0\farcs52$ (1 fibre) along the IFU long axis in order to assist cosmic ray rejection and facilitate masking of the dead fibres. 

Basic data reduction was carried out using the ESO GIRAFFE pipeline. This included bias subtraction, flat fielding, identification and spectrum extraction, and wavelength calibration. Post-processing was carried out in {\sc iraf}\footnote{The Image Reduction and Analysis Facility ({\sc iraf}) is distributed by the National Optical Astronomy Observatories which is operated by the Association of Universities for Research in Astronomy, Inc. under cooperative agreement with the National Science Foundation.}, including cosmic-ray rejection \citep[using \textsc{lacosmic};][]{vandokkum01}, flux calibration and the combining of the individual exposures (dead fibres were masked and removed when the dithered exposures were combined). The FWHM spectral resolution, measured from averaging fits to isolated arc lines over all spaxels, was found to be 0.55~\AA\ ($\sim$25~\kms\ at H$\alpha$).

\subsection{Emission line fitting} \label{sect:fitting}
The extremely high S/N (up to $\sim$20\,000 in H$\alpha$) and spectral resolution of these data have allowed us to quantify the emission line profile shapes to a high degree of accuracy. At the highest S/N ratios, an Airy diffraction pattern in the wings of the line profiles can be seen clearly. However, since it is only at the $\lesssim$1~percent level, it does not significantly interfere with our ability to fit the line profile shapes. We fitted multiple Gaussian profile models to the H$\alpha$, [N\two]$\lambda$6583 and [S\two]$\lambda$$\lambda$6717,6731 emission lines using an \textsc{idl}-based $\chi^{2}$ fitting package called \textsc{pan} \citep[Peak ANalysis;][]{azuah09}. Each line in each of the spaxels was fitted using a single-, double- and triple-Gaussian initial guess. Line fluxes were constrained to be positive and widths to be greater than the instrumental contribution to guard against spurious results. H$\alpha$, [N\two]$\lambda$6583, and [S\two]$\lambda$$\lambda$6717,6731 were fit independently due to the significantly different shapes of the lines. The [S\two] doublet was fitted simultaneously, where each component fit was constrained such that the wavelength difference between the two Gaussian models (one for $\lambda$6717 and one for $\lambda$6731) was equal to the laboratory difference, and FWHMs were equal to one another. Multi-component fits were run several times with different initial guess configurations (widths and wavelengths) in order to account for the varied profile shapes, and the result with the lowest $\chi^2$ was retained. However, we note that the $\chi^2$ minimisation routine employed by \textsc{pan} is very robust with respect to the initial guess parameters. 

To determine how many Gaussian components best fit an observed profile (one, two or three), we used a likelihood ratio test to determine whether a fit of $n$+1 components was more appropriate than an $n$-component fit. This test says that if the ratio of the $\chi^{2}$ statistics for the two fits falls above a certain threshold, then the fits are considered statistically distinguishable, and the one with the lower $\chi^{2}$ can be selected. Here we determine the threshold ratio by visual inspection of a range of spectra and fits. This method is a more generalised form of the formal F-test that we have used in previous work \citep[e.g.][]{westm09a, westm10b}. We are aware of work that cautions against the use of the F-test to test for the presence of a line \citep[or additional line component; e.g.][]{protassov02}, but given the absence of a statistically correct alternative that could be applied sensibly to the volume of data presented here, we have chosen to opt for this generalised $\chi^{2}$ ratio test approach.

This test, however, only tells us which of the fits (single, double or triple component) is most appropriate for the corresponding line profile. Experience has taught us that we need to apply a number of additional tests to filter out well-fit but physically improbable results. We specified that the fluxes of all components (in the fit selected by the likelihood ratio test) should be $>$0; we rejected fits where the FWHM of any component was more than 2.6~\AA\ since, from visual inspection, these tended to be where \textsc{pan} had attempted to fit the faint Airy rings; and we rejected any fits where $\chi^{2}_{\rm single}/\chi^{2}_{\rm double}=0$ (a symptom of a bad fit). We then applied a set of rules to assign each line component to a specific component set, which were chosen and refined after inspection of the resulting maps. For IFU Pos1, for a triple-Gaussian fit, we specified that the brightest component should be assigned to set C1, and after that, the redder to C2 and the bluer to C3. For double-Gaussian fits, we assigned C1 to be the bluer of the two and C2 to the redder. For IFU Pos2, no triple-component fits were required, and we assigned the brighter of any two-component fits to C1 and the fainter to C2. Producing consistent maps in this way helps limit the confusion that could arise during analysis of the results where discontinuous spatial regions might arise from incorrect component assignments.

\section{Emission line maps}

In the following, we present and discuss the integrated line flux maps, the individual component flux, FWHM, and radial velocity maps, and the line ratio maps produced using the emission line fits described above. We also show some example line profiles and their corresponding best-fitting multi-component models from a number of spaxels, as indicated on the maps with corresponding letters. Below each plot are the corresponding residuals, $r_{\rm i}$, calculated using the following formula:
\begin{equation}
r_{\rm i} = \frac{y_{\rm i}^{\rm fit} - y_{\rm i}^{\rm data}}{\sigma_{\rm i}}
\end{equation}
where $\sigma_{\rm i}$ are the uncertainties on $y_{\rm i}^{\rm data}$. However, since the uncertainties are not accurately quantified (uncertainties are not a product in the current reduction pipeline), it is only the shape of the residual plot that has any meaning. In some plots (e.g.\ Fig.~\ref{fig:Ha_IFU1_egfits}d), the aforementioned Airy rings are evident, particularly in the residual plot.

\subsection{Position 1} \label{sect:ARGUS_maps1}
Position 1 covers the very tip of a pillar located to the west of the star cluster and an emission structure immediately to the east of the pillar, including an interesting arc of ionized gas that is prominent in H$\alpha$. Examination of the morphology, kinematics and line ratios allow us to determine the state of the ionized gas in these structures and whether they are associated with each other or not.

\begin{figure*}
\begin{minipage}{5.5cm}
\includegraphics[width=\textwidth]{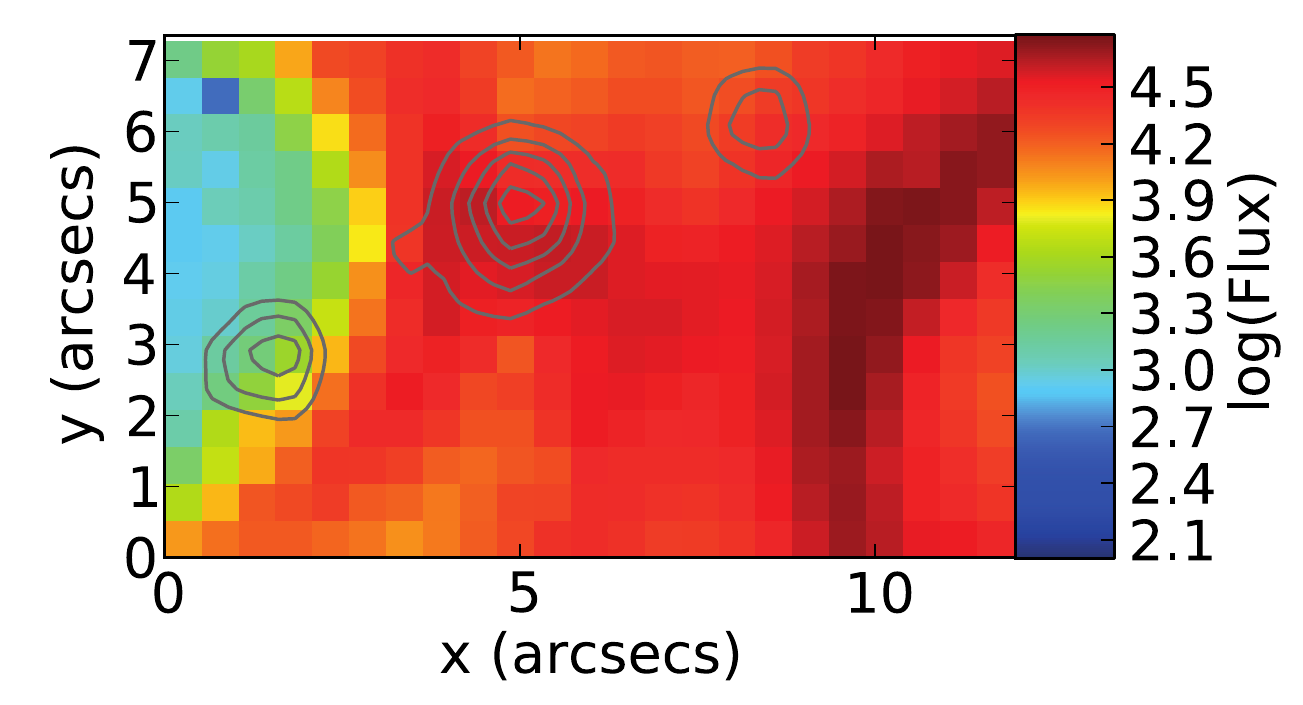}
\end{minipage}
\begin{minipage}{5.5cm}
\includegraphics[width=\textwidth]{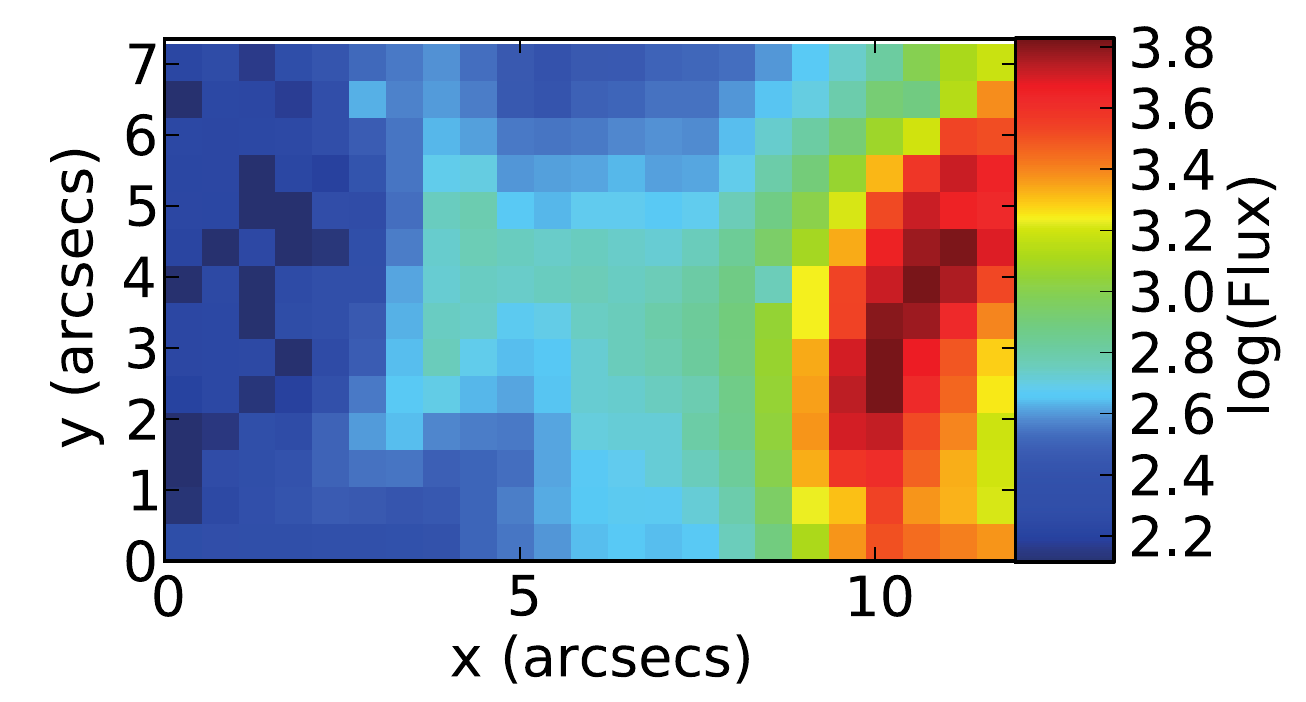}
\end{minipage}
\begin{minipage}{5.5cm}
\includegraphics[width=\textwidth]{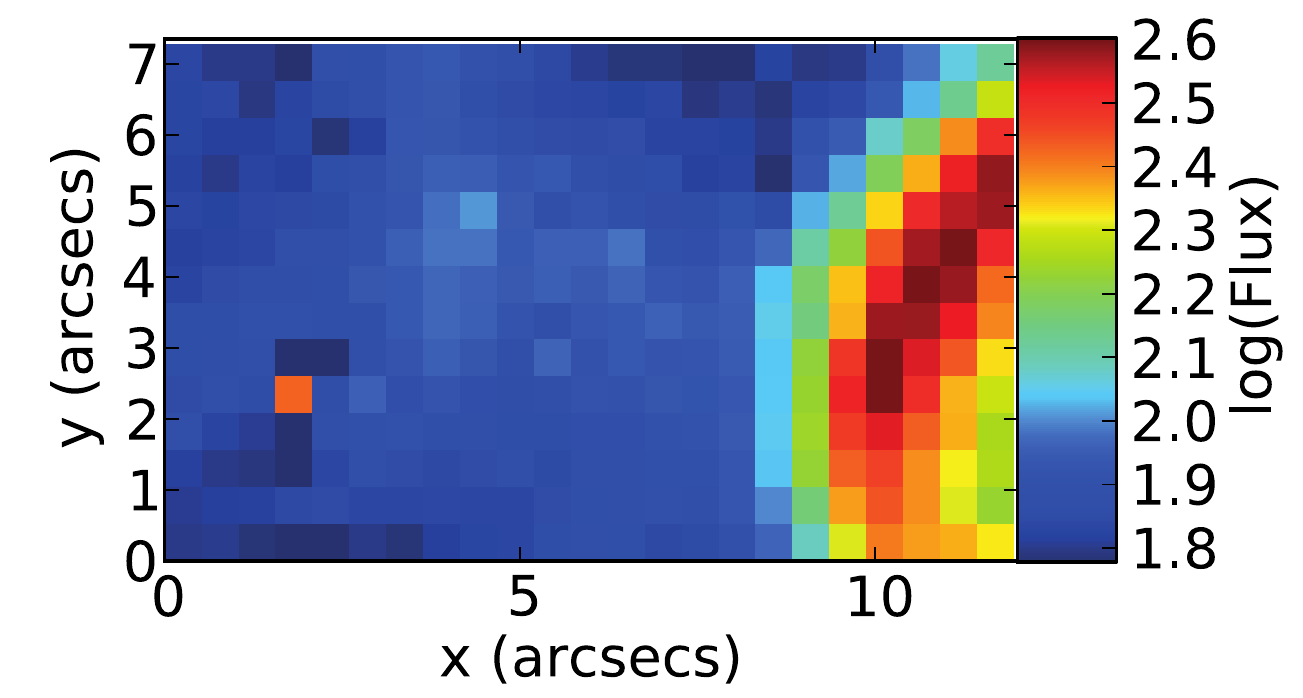}
\end{minipage}
\caption{H$\alpha$ (left), [N\two] (centre) and [S\two] (right) summed intensity maps scaled to the maximum/minimum in each map. Contours on the H$\alpha$ map represent the continuum level, and highlight the location of three stars in the field-of-view.}
\label{fig:sum_IFU1}
\end{figure*}

Fig.~\ref{fig:sum_IFU1} shows maps of the integrated line flux (C1+C2+C3) for H$\alpha$, [N\two]$\lambda$6583 and [S\two]$\lambda$6717. The surface of the pillar tip, to the west of the field-of-view (FoV), is clearly visible in all three maps, whereas the arc-shaped feature towards the east is much more prominent in H$\alpha$ (and not visible at all in [S\two]). The spatially-resolved line ratio maps will be discussed below in Section~\ref{IFU1_ratios}.

\begin{figure*}
\begin{minipage}{15cm}
\begin{overpic}[width=\textwidth]{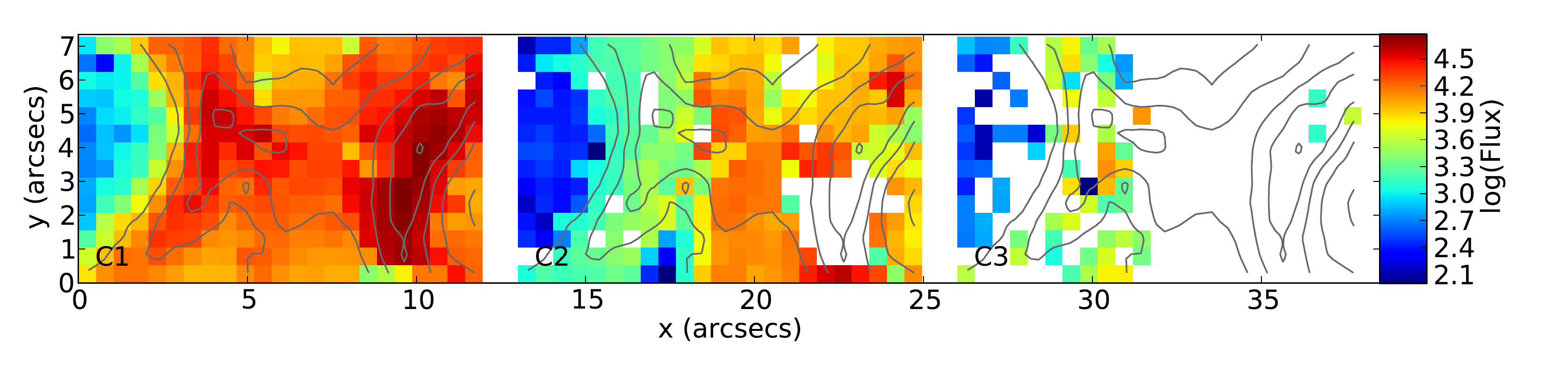}
\put(6,13){\circled{\Large{\textcolor{white}{a}}}}
\put(24,10){\circled{\Large{\textcolor{white}{d}}}}
\put(12,16){\circled{\Large{\textcolor{white}{b}}}}
\put(25,18){\circled{\Large{\textcolor{white}{e}}}}
\put(17,9){\circled{\Large{\textcolor{white}{c}}}}
\end{overpic}
\end{minipage}
\begin{minipage}{15cm}
\includegraphics[width=\textwidth]{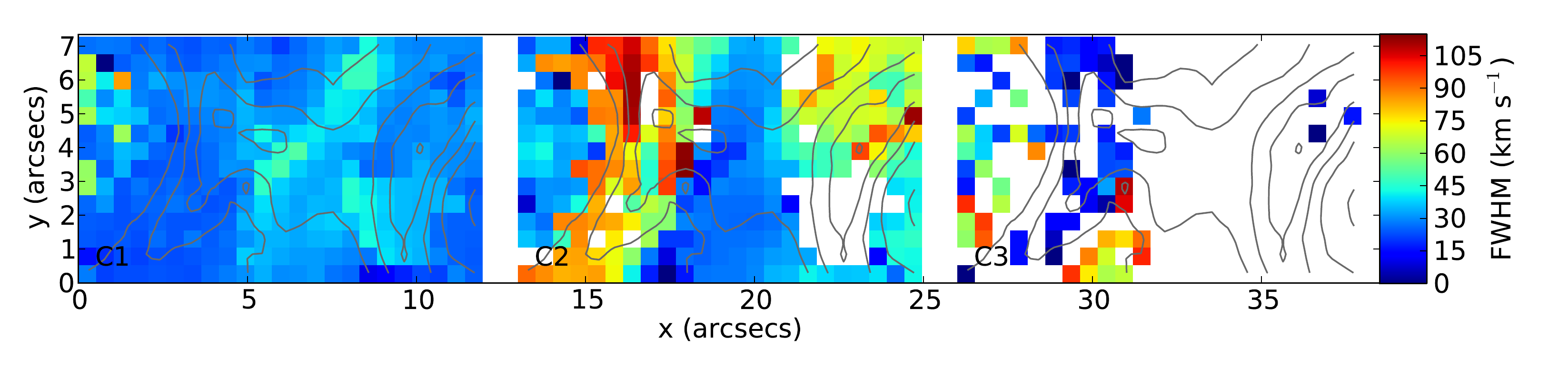}
\end{minipage}
\begin{minipage}{15cm}
\includegraphics[width=\textwidth]{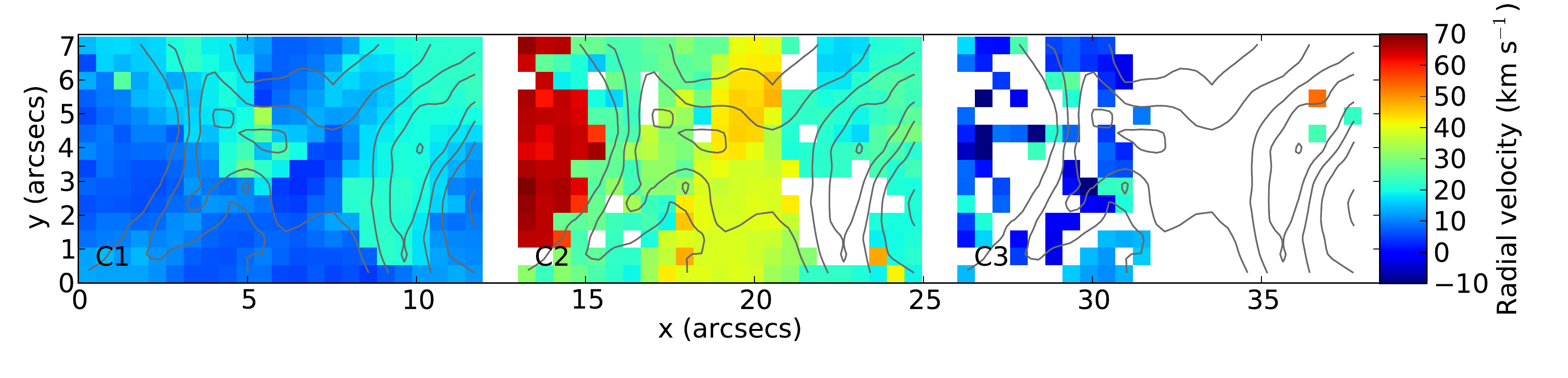}
\end{minipage}
\caption{H$\alpha$ line component maps for IFU position 1. Component 1 (C1) is shown on the left, component 2 (C2) in the centre (starting at $x=13''$), and component 3 (C3) on the right (starting at $x=26''$). The flux scale bar is in units of $10^{-18}$~erg~s$^{-1}$~cm$^{-2}$~spaxel$^{-1}$, the line widths are corrected in quadrature for the contribution of instrumental broadening, and the velocity maps are in the heliocentric frame of reference.  Grey contours in this and the subsequent maps represent the total (C1+C2+C3) H$\alpha$ flux. The spaxels from which the example line profiles shown in Fig.~\ref{fig:Ha_IFU2_egfits} were extracted are labelled with the corresponding letters.}
\label{fig:Ha_IFU1}
\end{figure*}

\begin{figure*}
\begin{minipage}{15cm}
\begin{overpic}[width=\textwidth]{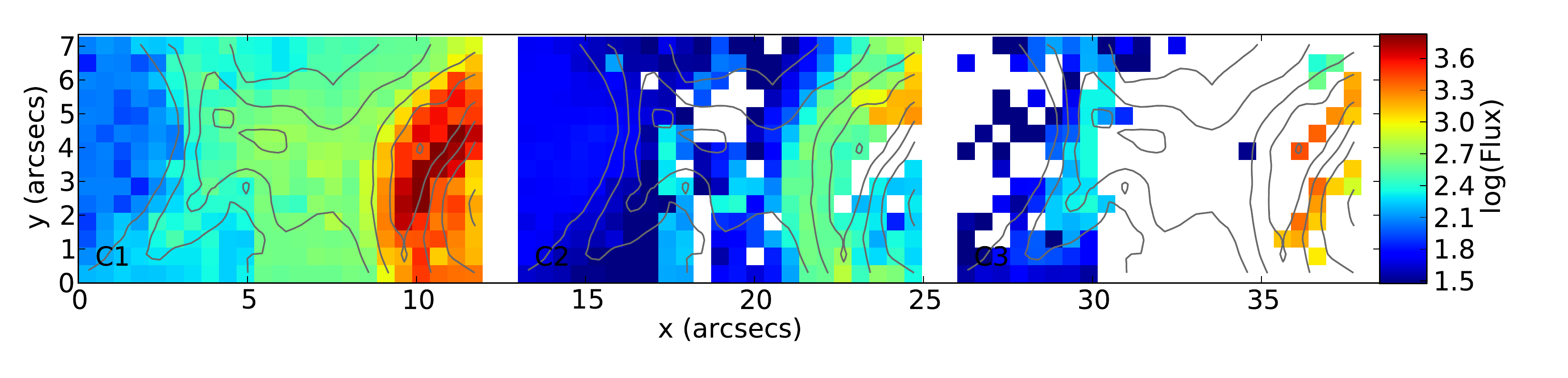}
\put(11,15){\circled{\Large{\textcolor{white}{a}}}}
\put(25,14){\circled{\Large{\textcolor{white}{b}}}}
\end{overpic}
\end{minipage}
\begin{minipage}{15cm}
\vspace*{-0.3cm}
\includegraphics[width=\textwidth]{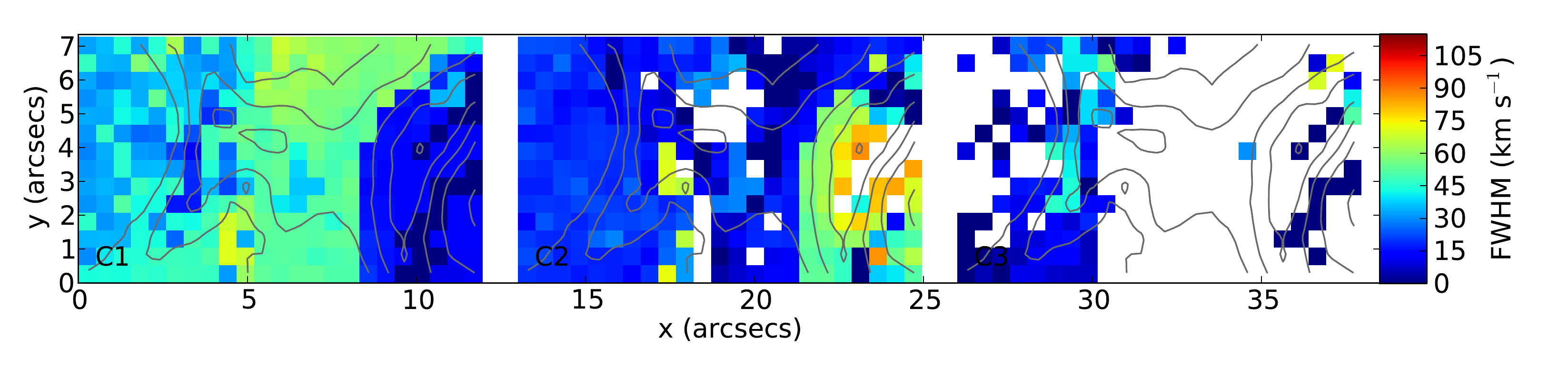}
\end{minipage}
\begin{minipage}{15cm}
\vspace*{-0.3cm}
\includegraphics[width=\textwidth]{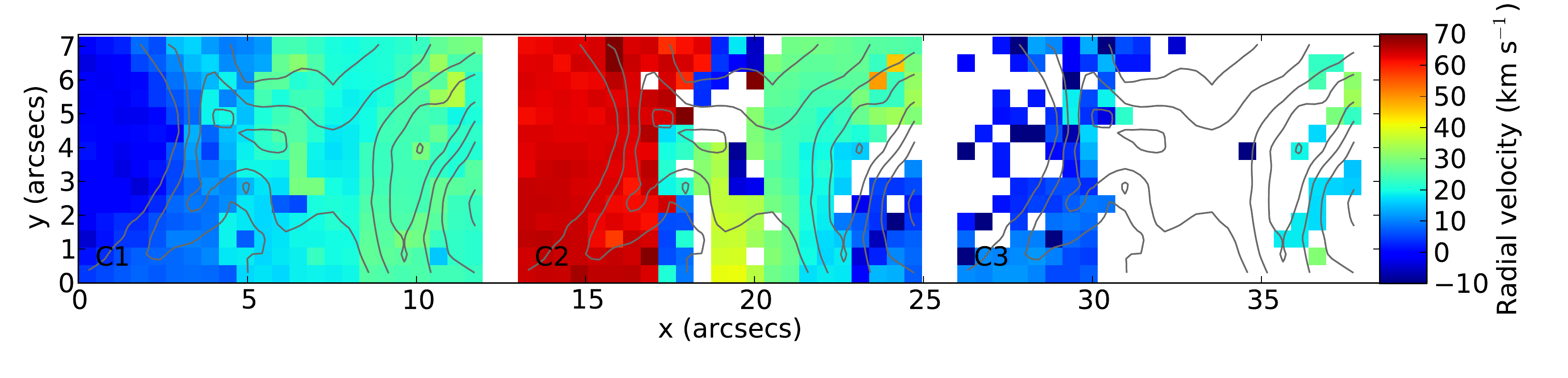}
\end{minipage}
\caption{As Fig.~\ref{fig:Ha_IFU1} but for the [N\two] line components.}
\label{fig:NII_IFU1}
\end{figure*}

Maps of the flux, FWHM, and radial velocity for the individual line components in H$\alpha$ and [N\two] are shown in Figs.~\ref{fig:Ha_IFU1} and \ref{fig:NII_IFU1}, respectively. The line widths are corrected for the contribution of instrumental broadening, and the velocity maps are in the heliocentric frame of reference. On each map, contours representing the integrated H$\alpha$ flux distribution from Fig.~\ref{fig:sum_IFU1} are shown to locate the tip of the pillar and other emission features, and help guide the eye. Figs.~\ref{fig:Ha_IFU1_egfits} and \ref{fig:NII_IFU1_egfits} show example line profile fits for H$\alpha$ and [N\two], respectively, chosen to illustrate the variety of line shapes found and the quality of the line fits we were able to achieve.

\subsubsection{H$\alpha$ maps}

\begin{figure*}
\begin{minipage}{5.2cm}
\begin{overpic}[width=\textwidth]{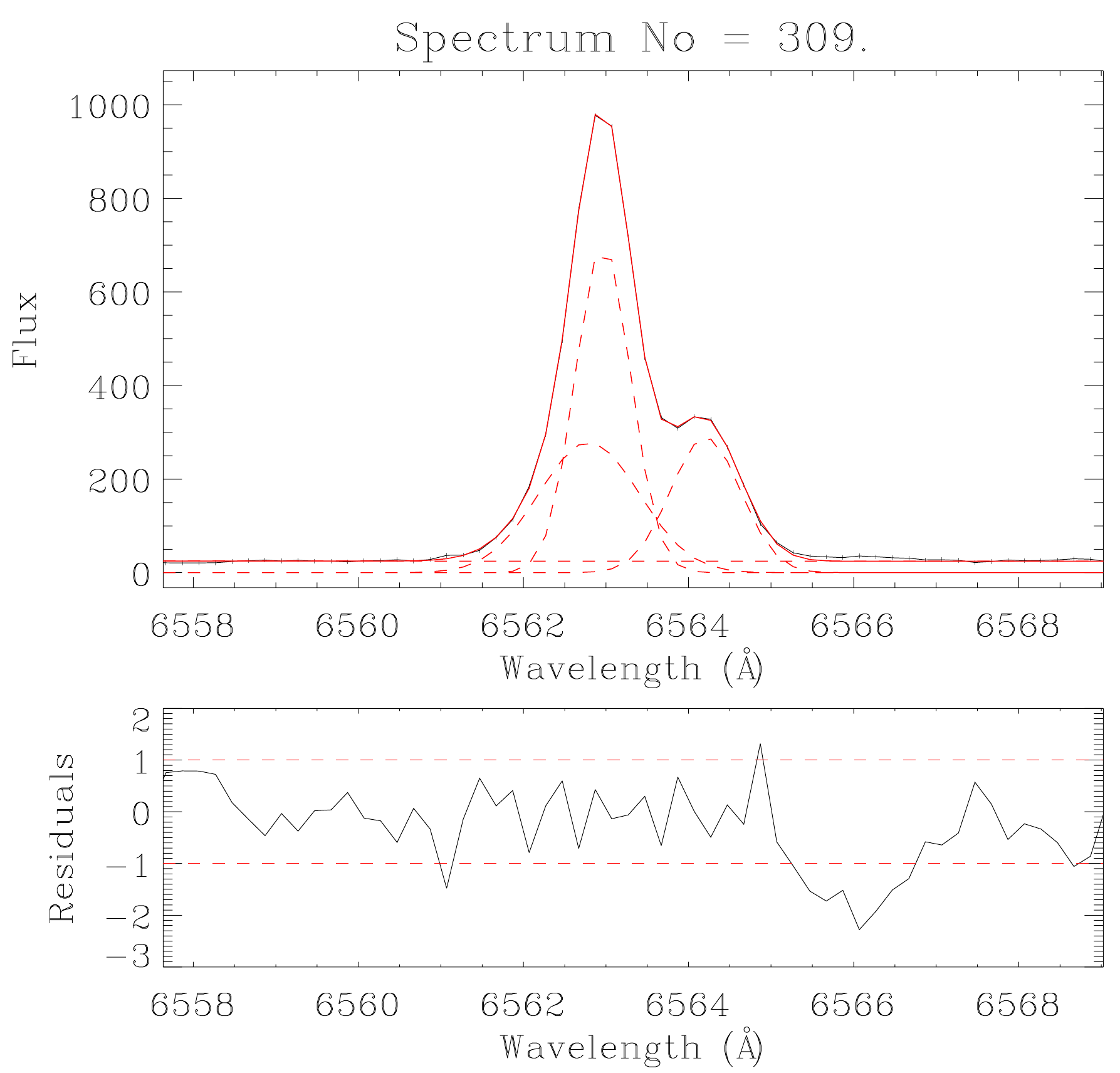}
\put(21,80){(a)}
\end{overpic}
\end{minipage}
\begin{minipage}{5cm}
\begin{overpic}[width=\textwidth]{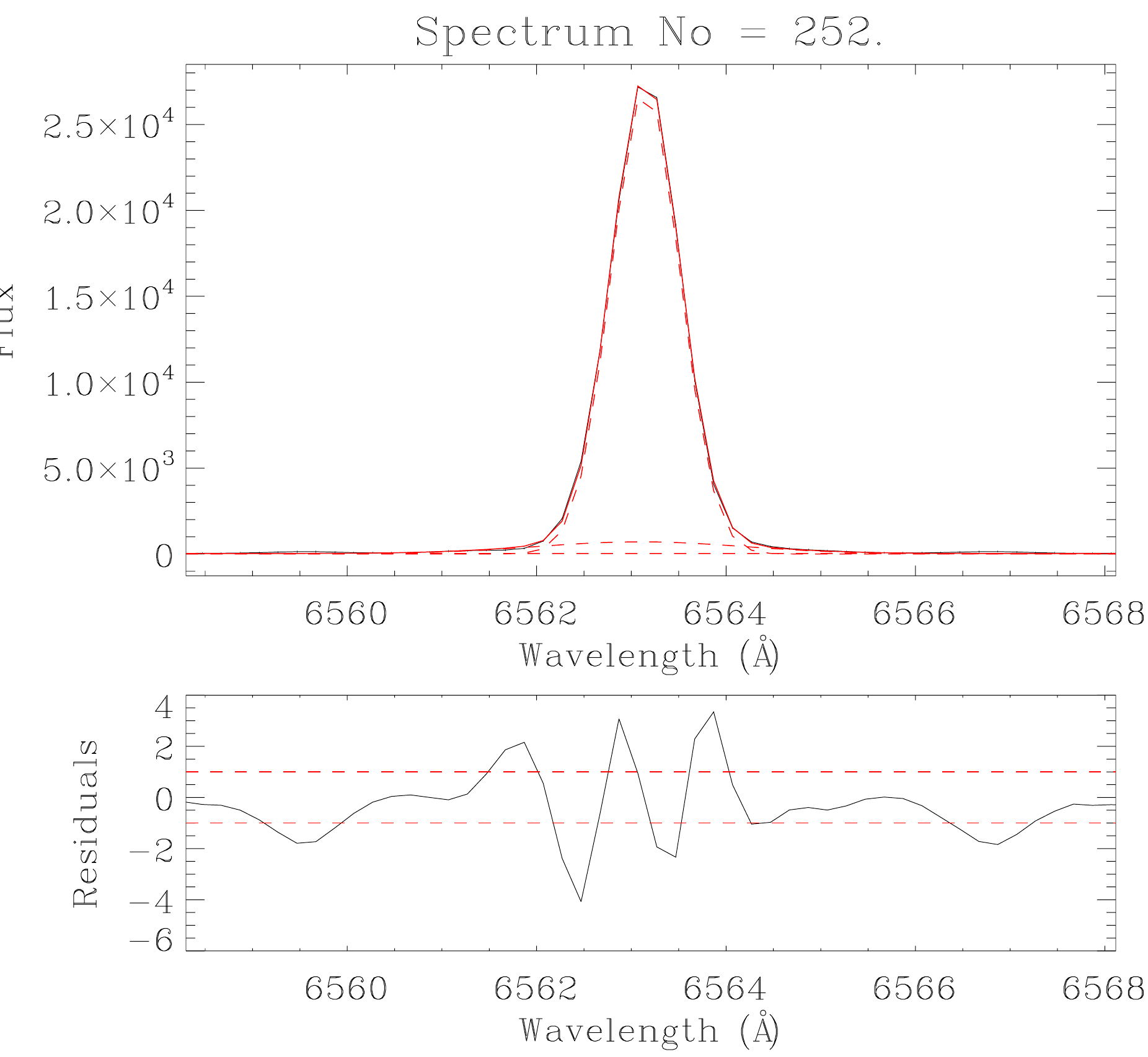}
\put(21,80){(b)}
\end{overpic}
\end{minipage}
\begin{minipage}{5cm}
\begin{overpic}[width=\textwidth]{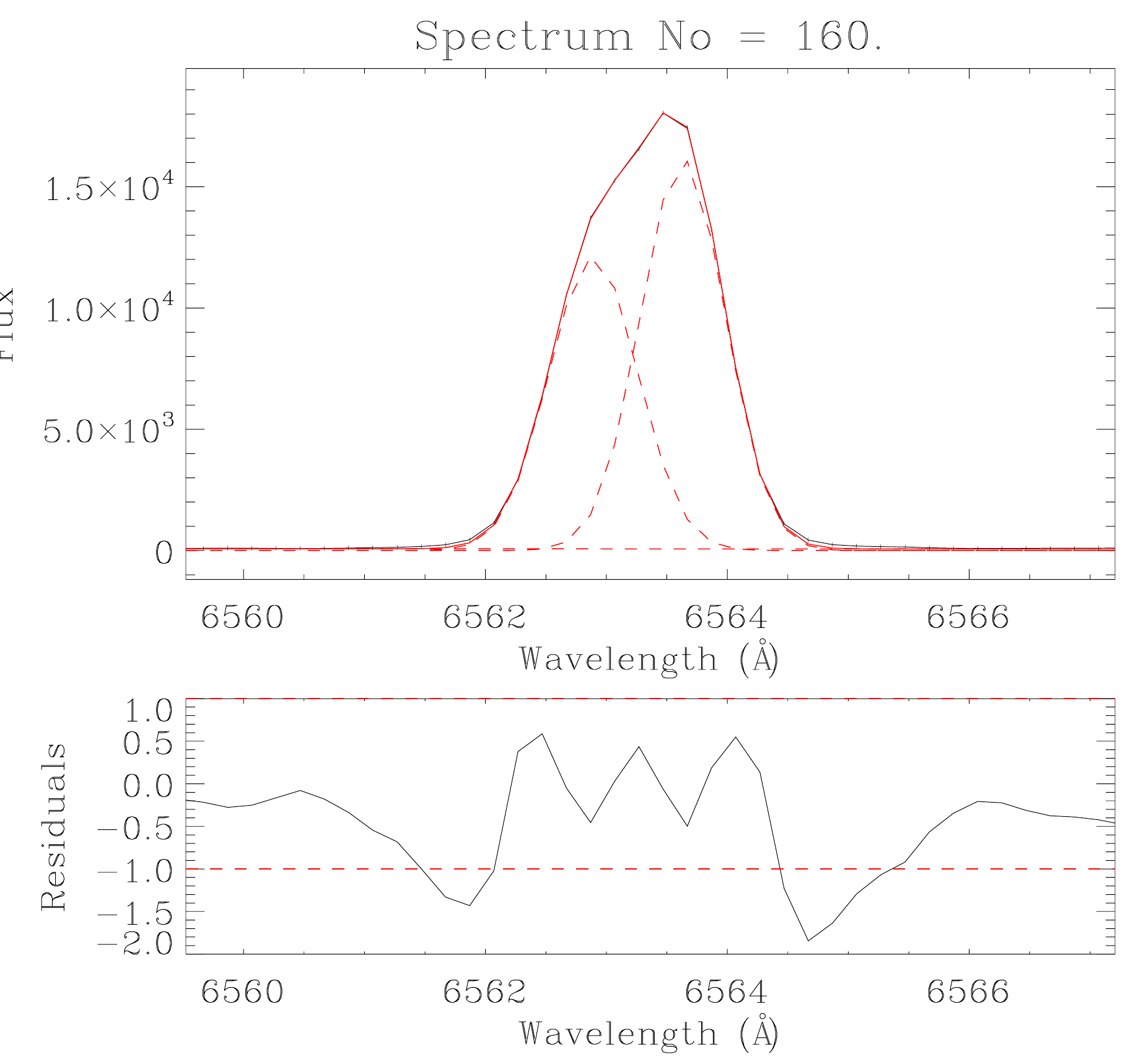}
\put(21,80){(c)}
\end{overpic}
\end{minipage}
\begin{minipage}{5cm}
\begin{overpic}[width=\textwidth]{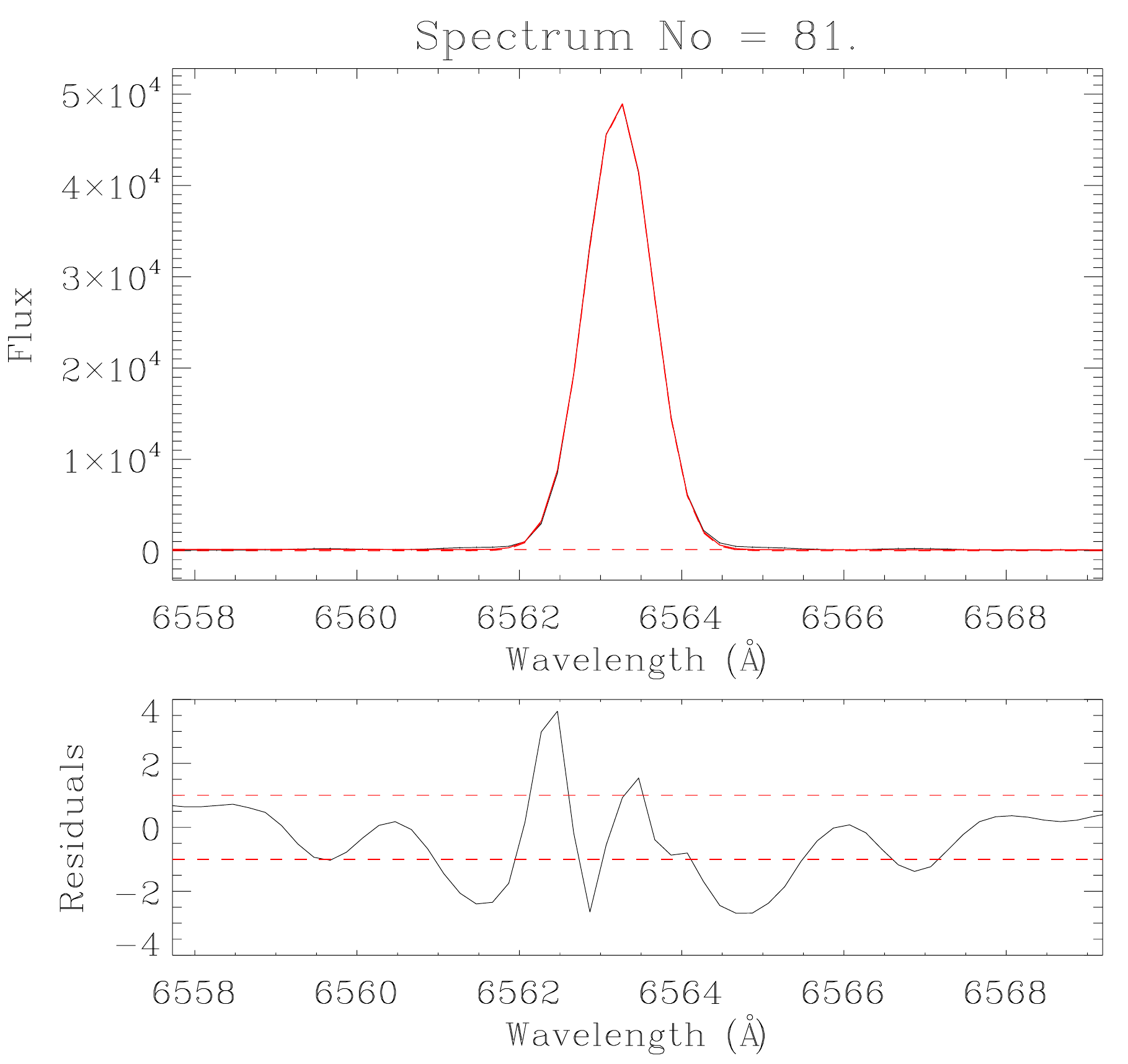}
\put(21,80){(d)}
\end{overpic}
\end{minipage}
\begin{minipage}{5cm}
\begin{overpic}[width=\textwidth]{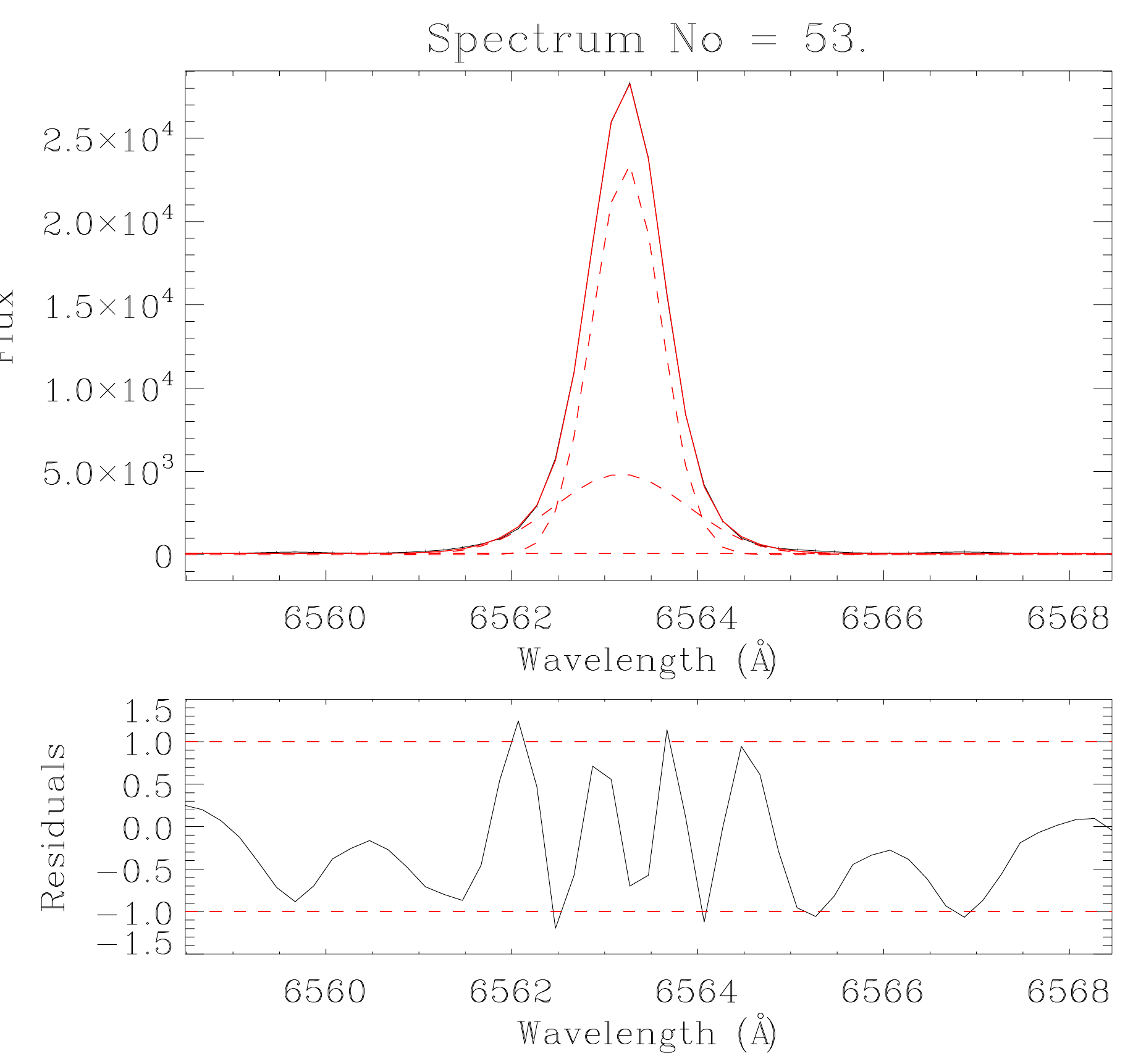}
\put(21,80){(e)}
\end{overpic}
\end{minipage}
\caption{Example H$\alpha$ line profiles extracted from the locations shown in Fig.~\ref{fig:Ha_IFU1}, chosen to represent the main types of profile shapes observed over the field. Observed data are shown by a solid black line, individual Gaussian fits by dashed red lines (including the straight-line continuum level fit), and the summed model profile in solid red. Flux units are $10^{-18}$~erg~s$^{-1}$~cm$^{-2}$~\AA$^{-1}$~spaxel$^{-1}$. Below each spectrum is the residual plot.}
\label{fig:Ha_IFU1_egfits}
\end{figure*}

C1 is the brightest component across the whole field (recall that by definition it is the bluer of any double-component fits). Its width is consistently narrow (20--40~\kms) and it shows very little velocity structure. The pillar tip is slightly redshifted compared to the rest of the field (by $\sim$10--15~\kms), and is certainly the brightest H$\alpha$-emitting region. The northern edge of the pillar tip exhibits a profile shape that is best fit with a bright, narrow ($\sim$30~\kms) component and and underlying broad ($\sim$70~\kms), fainter component (line fit example Fig.~\ref{fig:Ha_IFU1_egfits}e). 
A similar line shape (narrow + faint broad) is found in the arc towards the east, and here the broad component has widths of 90--110~\kms\ (line fit example Fig.~\ref{fig:Ha_IFU1_egfits}b). In the region between the arc and the pillar head, a secondary narrow component is identified that has a velocity $\sim$20--25~\kms\ redshifted with respect to the pillar and 30--40~\kms\ redshifted compared to C1 at the same location (example Fig.~\ref{fig:Ha_IFU1_egfits}c). The semi-circular region to east of the arc at the far eastern edge of the FoV, is, in comparison, faint in emission compared to the rest of the field. Fig.~\ref{fig:Ha_IFU1_egfits}a shows an example line profile from this region. Here, two main components in H$\alpha$ are identified, one at a velocity similar to pillar surroundings ($\sim$5~\kms), and one much fainter but redshifted by 60--70~\kms\ (note that the S/N ratio of this fainter component is still $\sim$160). That this redshifted component is not visible anywhere else in the field suggests that it originates from gas located in the background and is obscured by the gas structures in the rest of the field. A third component, narrow and blueshifted (by $\sim$20~\kms\ with respect to the pillar), can be identified over the whole eastern half of field.

\subsubsection{[N\two] maps}
\begin{figure*}
\begin{minipage}{5.2cm}
\begin{overpic}[width=0.98\textwidth]{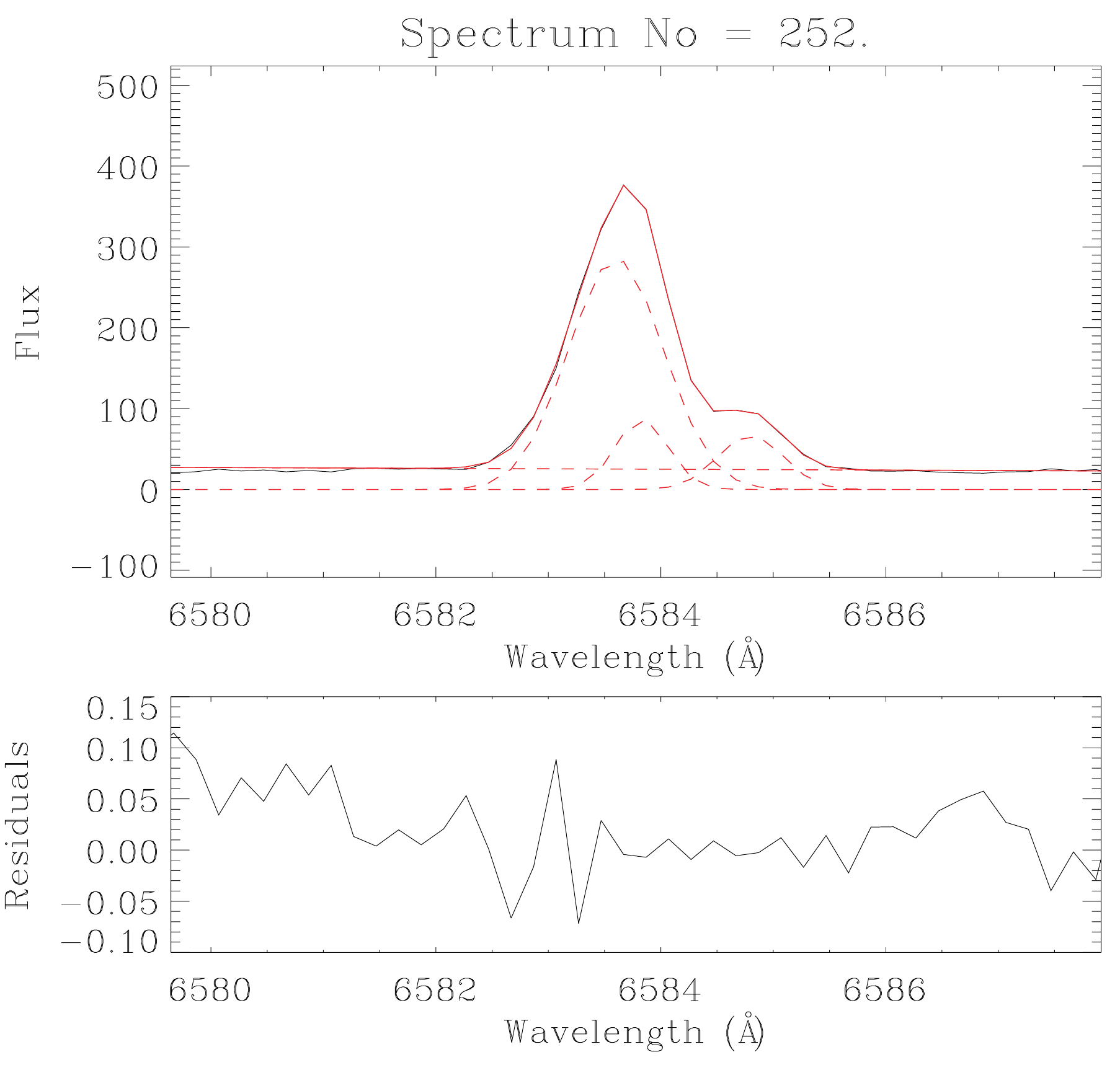}
\put(21,80){(a)}
\end{overpic}
\end{minipage}
\begin{minipage}{5cm}
\begin{overpic}[width=\textwidth]{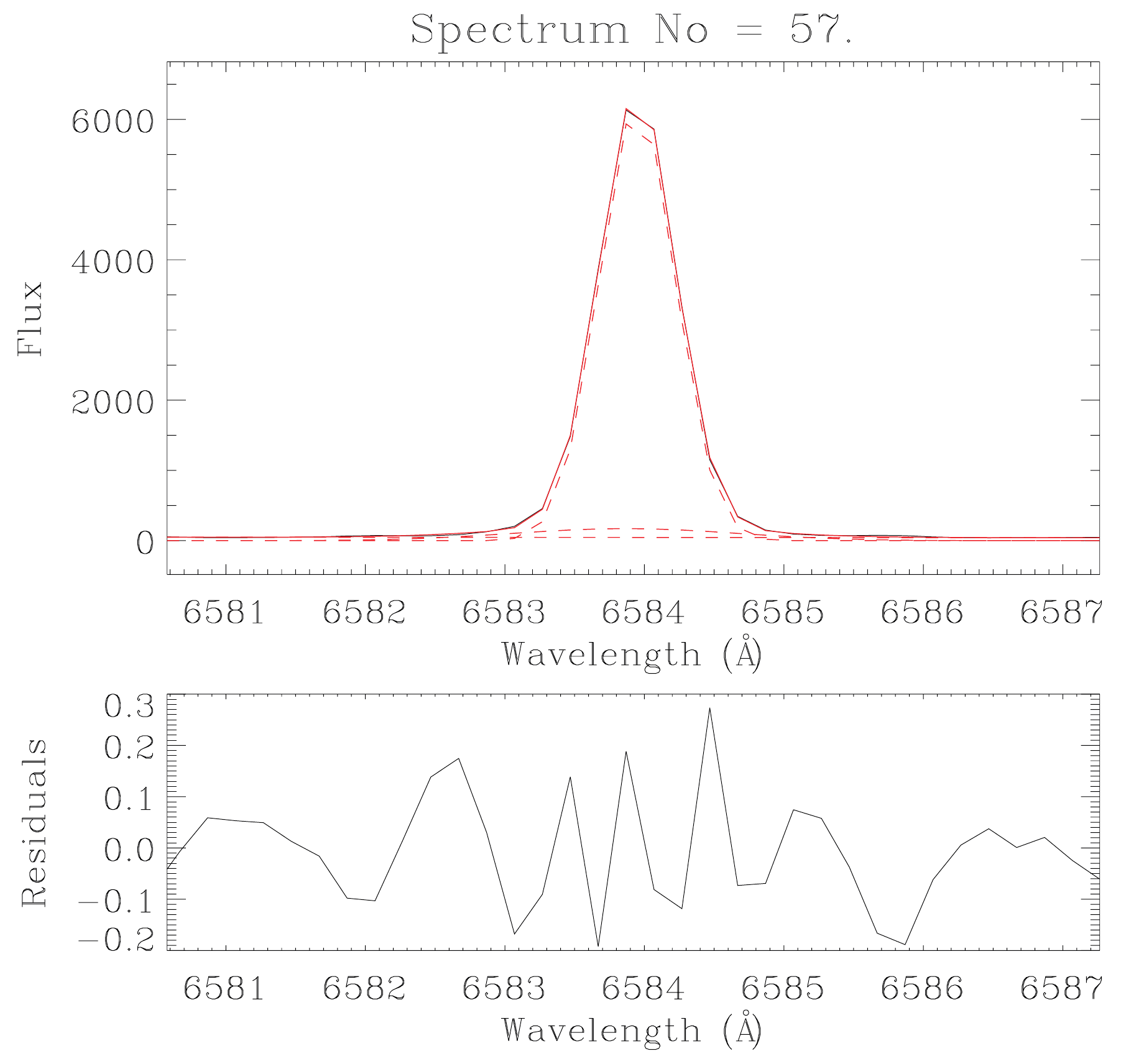}
\put(21,80){(b)}
\end{overpic}
\end{minipage}
\caption{Example [N\two] line profiles extracted from the locations shown in Fig.~\ref{fig:NII_IFU1}. See the caption to Fig.~\ref{fig:Ha_IFU1_egfits} for further details.}
\label{fig:NII_IFU1_egfits}
\end{figure*}

The emission profile shapes of the [N\two] line are broadly similar to that of H$\alpha$ (recall that we fit each line independently). However, as mentioned above, the arc and filamentary emission to the east of the field is much less prominent. The faint 60--70~\kms\ redshifted component, seen in H$\alpha$ only in the far east of the field (to the east of the arc) is visible in [N\two] much further towards the pillar (as demonstrated in the example line fit Fig.~\ref{fig:NII_IFU1_egfits}a). The weakness of the [N\two] emission from the arc must allow this background component to be visible where in H$\alpha$ it is not. The broad C2 component seen in H$\alpha$ on the arc is not found in [N\two], although it is on the pillar tip (FWHM$\sim$70--80~\kms; example Fig.~\ref{fig:NII_IFU1_egfits}b).

The [S\two] line maps are very similar to the [N\two] maps; due to their lower S/N we do not show them here.

\subsubsection{Line ratios} \label{IFU1_ratios}
The forbidden/recombination line flux ratios of [N\two]$\lambda$6583/H$\alpha$ and [S\two]($\lambda$6717+$\lambda$6731)/H$\alpha$ can be used as indicators of the number of ionizations per unit volume \citep{veilleux87,dopita95,dopita06b}. Both [N\two]/H$\alpha$ and [S\two]/H$\alpha$ are also sensitive to shock ionization since relatively cool, high-density regions form behind shock fronts which emit strongly in these forbidden lines, resulting in an enhancement in the forbidden/recombination line ratios \citep{dopita97, oey00}. These indicators (normally combined with a tracer of higher energy ionizing photons such as [O\three]/H$\beta$; i.e.\ a BPT diagram; \citealt{baldwin81}) can be used to establish the dominance of non-photoionization processes (i.e.\ shocks -- or AGN, but that is not relevant here) via a threshold ratio \citep[theoretically calculated;][]{dopita95, kewley01, calzetti04} or by comparison to shock models \citep[e.g.][]{allen08}. However, a certain amount of caution is needed when applying these classical diagnostic diagrams and theoretical ratios derived from 1D photoionization calculations to detailed spatially resolved observations of complex 3D star-forming regions \citep{ercolano12}.

\begin{figure*}
\begin{minipage}{15cm}
\includegraphics[width=\textwidth]{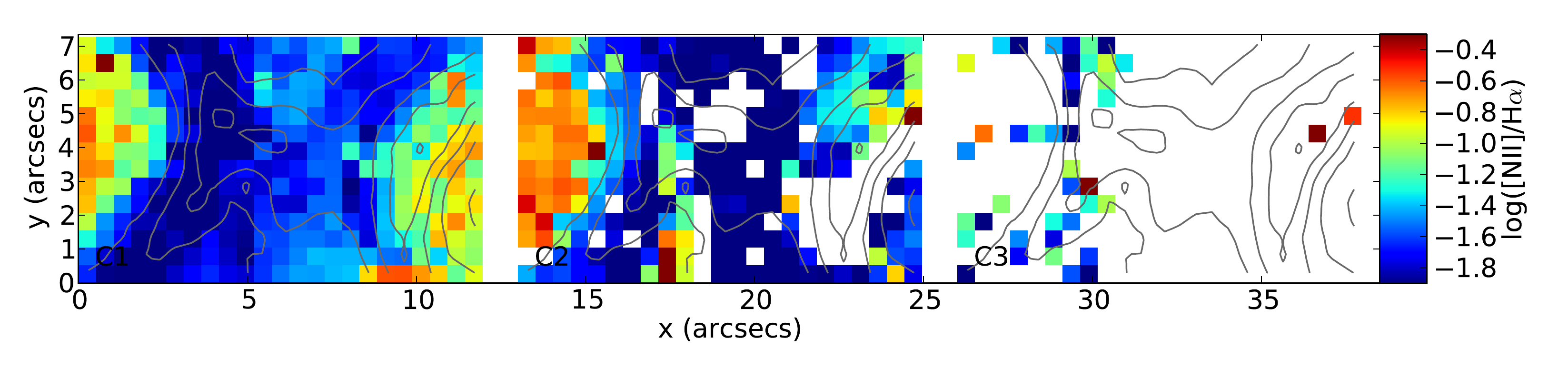}
\end{minipage}
\begin{minipage}{15cm}
\includegraphics[width=\textwidth]{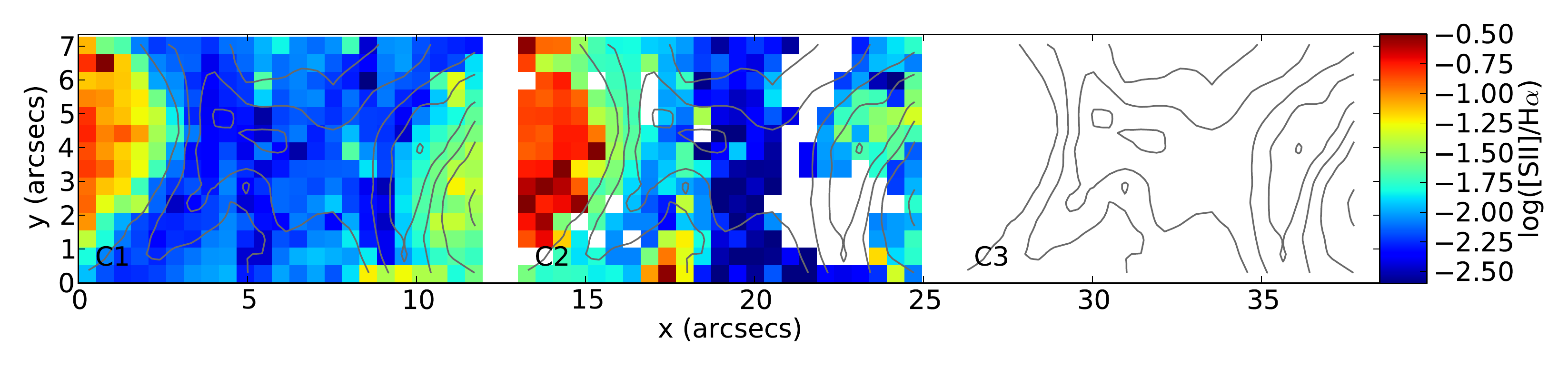}
\end{minipage}
\caption{[N\two]/H$\alpha$ and [S\two]/H$\alpha$ line ratio maps for each component.}
\label{fig:ratios_IFU1}
\end{figure*}

Maps of the [N\two]/H$\alpha$ and [S\two]/H$\alpha$ ratios in the three line components identified are shown in Fig.~\ref{fig:ratios_IFU1}. Line ratios were only calculated where the corresponding components in both (or all three of) the lines were identified. A third component of [S\two] could not be identified in any spaxel, hence the map is blank. The maximum line ratios in both components C1 and C2 are found in the semi-circular region in the far eastern side of the field, which from the kinematical maps we suggest is in the far background, behind the pillar and arc. In C1, both ratios are slightly lower on the pillar tip, and significantly lower in the intervening gas, including the arc. This is mirrored in C2, but the lack of a detection of this component in places makes it less clear. The arc, which exhibits such broad line components in H$\alpha$, is not prominent in these maps.

[N\two]/H$\alpha$ and [S\two]/H$\alpha$ ratios greater than $-0.1$ and $-0.4$, respectively, would indicate a strong contribution from shocks to the ionization budget \citep{allen08}. However at no point are the line ratios we measure greater than these values. Nevertheless, the motions of the turbulent gas are clearly supersonic (the sound speed in $10^4$~K gas is $\sim$10~\kms) meaning that shocks are undoubtedly present. It is possible the reason why we do not see signatures of shocks (i.e.\ enhanced forbidden line emission from the post-shock gas) is that in these small spatially resolved regions the shock contribution is completely washed out by contribution from photoionization. Shocks must be present, but we simply cannot see them in the line ratios.

\begin{figure*}
\includegraphics[width=15cm]{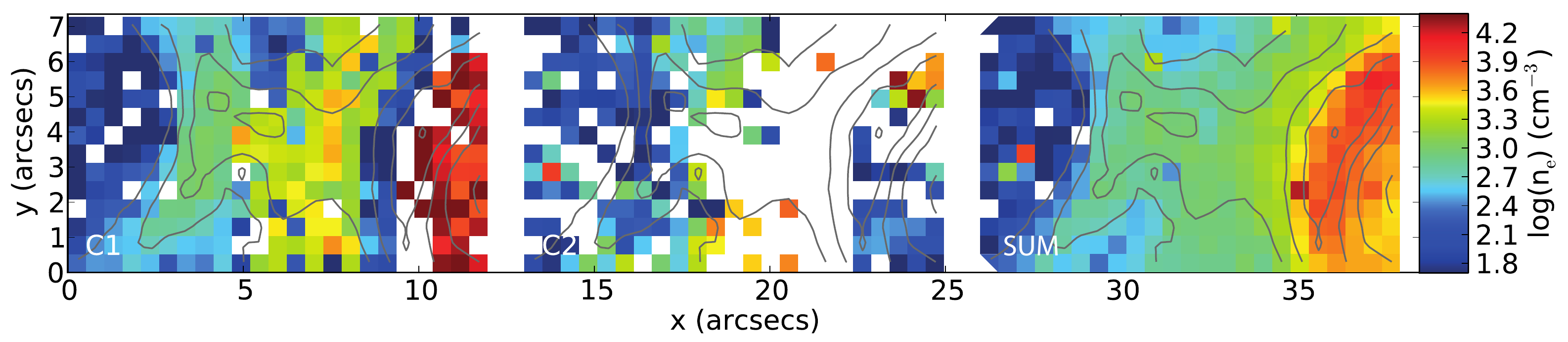}
\caption{Electron density, $n_{\rm e}$, maps derived from the C1 (left), C2 (centre) and summed (C1+C2; right) line flux ratios of the [S\two]$\lambda$$\lambda$6717,6731 doublet. Missing spaxels show where either the [S\two] component was undetected or the density calculation failed due to an unphysical ratio (a symptom of a bad fit due to low S/N). In some spaxels the density calculation from the individual component results fails, where it does not for the summed values.}
\label{fig:IFU1_elecdens}
\end{figure*}

Fig.~\ref{fig:IFU1_elecdens} shows the electron density map derived from the C1, C2 and summed (C1+C2) line flux ratios of the [S\two] doublet. The electron density of the semi-circular region of unobscured background gas at the east of the field is at or below the low-density limit for this indicator (50--100~\cmt). However, the material in the foreground is at significantly higher densities: the arc and gas between the arc and pillar has densities $>$500~\cmt, up to a few 1000~\cmt, and the pillar itself has electron densities over 10\,000~\cmt\ (close to the high-density limit for this [S\two] doublet method). The individual component maps show that the high densities of the arc and pillar are only seen in C1, whereas the equivalent C2 densities are much lower. The high C1 densities on the pillar imply that the majority of the narrow component [S\two] emission here originates from gas near to the (deeper, denser) neutral/molecular layers of the pillar that is just being ionized. 

\citet{pound98} estimated the molecular $n$(H$_{2}$) density of the pillar tips in the Eagle Nebula (M16) to be $\sim$$2\times10^5$~\cmt, while \citet{pound03} found an average $n$(H$_{2}$) of 4800~\cmt\ in the Horsehead Nebula, and \citet{gahm06} found $n$(H$_{2}$) densities of $10^3$--$10^4$~\cmt\ in a sample of nearby pillars. Thus, since this pillar exhibits ionized gas densities that are already comparable to these molecular gas densities, the core of this pillar must be particularly dense.

\subsubsection{Summary}
In H$\alpha$ we find broad (FWHM=90--110~\kms) line components associated with the arc, and broad (FWHM$\sim$70~\kms) components partly on the pillar tip. The fact that the arc also exhibits broad line widths implies that either it is also experiencing an impact of the star cluster wind, or that it is a highly turbulent eddy in the material being swept up by the cluster wind. The centroid radial velocities of the broad components on the arc and pillar are consistent with their corresponding narrow components, suggesting that they originate in gas at the same location. Furthermore, the radial velocities of the arc and the pillar are similar, implying that they lie on the same plane along the line of sight. In [N\two] we find a broad component (70--80~\kms) on the pillar tip but not at the location of the arc, and the [N\two]/H$\alpha$ and [S\two]/H$\alpha$ line ratios on the pillar are enhanced compared to the surrounding gas. The narrow-component electron densities on the pillar are surprisingly high ($>$10\,000~\cmt) compared to the arc and gas surrounding the pillar tip ($>$500~\cmt, up to a few 1000~\cmt), and that of the background gas (at or below 100~\cmt). The corresponding broad-component densities are much lower, implying that the narrow component originates from gas near to the (deeper, denser) neutral/molecular layers of the pillar (that is just being ionized) whereas the broad component originates in material further out.

At the far eastern edge of the FoV we sample a region of gas that exhibits a double narrow component line shape in both H$\alpha$ and [N\two], the fainter of which (C2) is redshifted. We conclude that this component originates from gas in the background behind the arc and pillar. In [N\two], the region in which we observe this redshifted component extends further into the IFU field towards the pillar. This background dynamical component exhibits high [N\two]/H$\alpha$ and [S\two]/H$\alpha$ ratios (although C1 in the same location also exhibits high line ratios), suggesting that the ambient gas behind the pillar has a high excitation (and, from above, low density). This is consistent with being the rarified, high excitation gas expected to exist in the hot bubble interior.

Between the arc and the pillar we find a narrow H$\alpha$ component that is redshifted compared to the arc and pillar, but not as much as the gas at the far east. This suggests we are seeing three planes along the line of sight: the arc and pillar in the foreground, the filamentary material between the two in the middle distance, and the redshifted gas to the east of the arc in the far background. The filamentary material in the middle-distance is likely part of the ionization/shock front SF1 identified by \citet{nurnberger03}, which extends another $\sim$100$''$ ($\sim$3.3~pc) to the north-west.

%
\subsection{Position 2} \label{sect:maps_IFU2}
\begin{figure*}
\begin{minipage}{5.2cm}
\includegraphics[width=\textwidth]{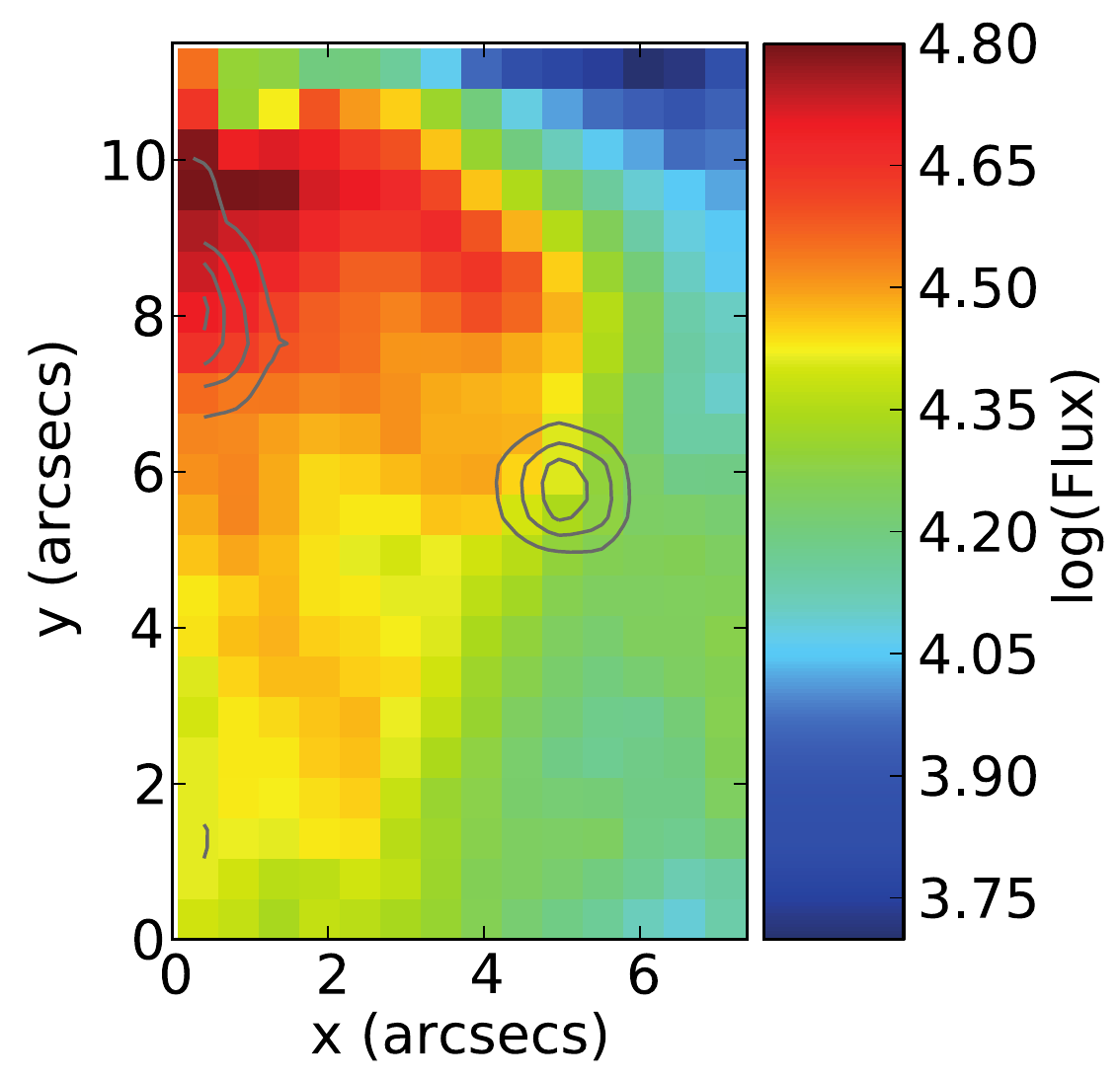}
\end{minipage}
\begin{minipage}{5cm}
\includegraphics[width=\textwidth]{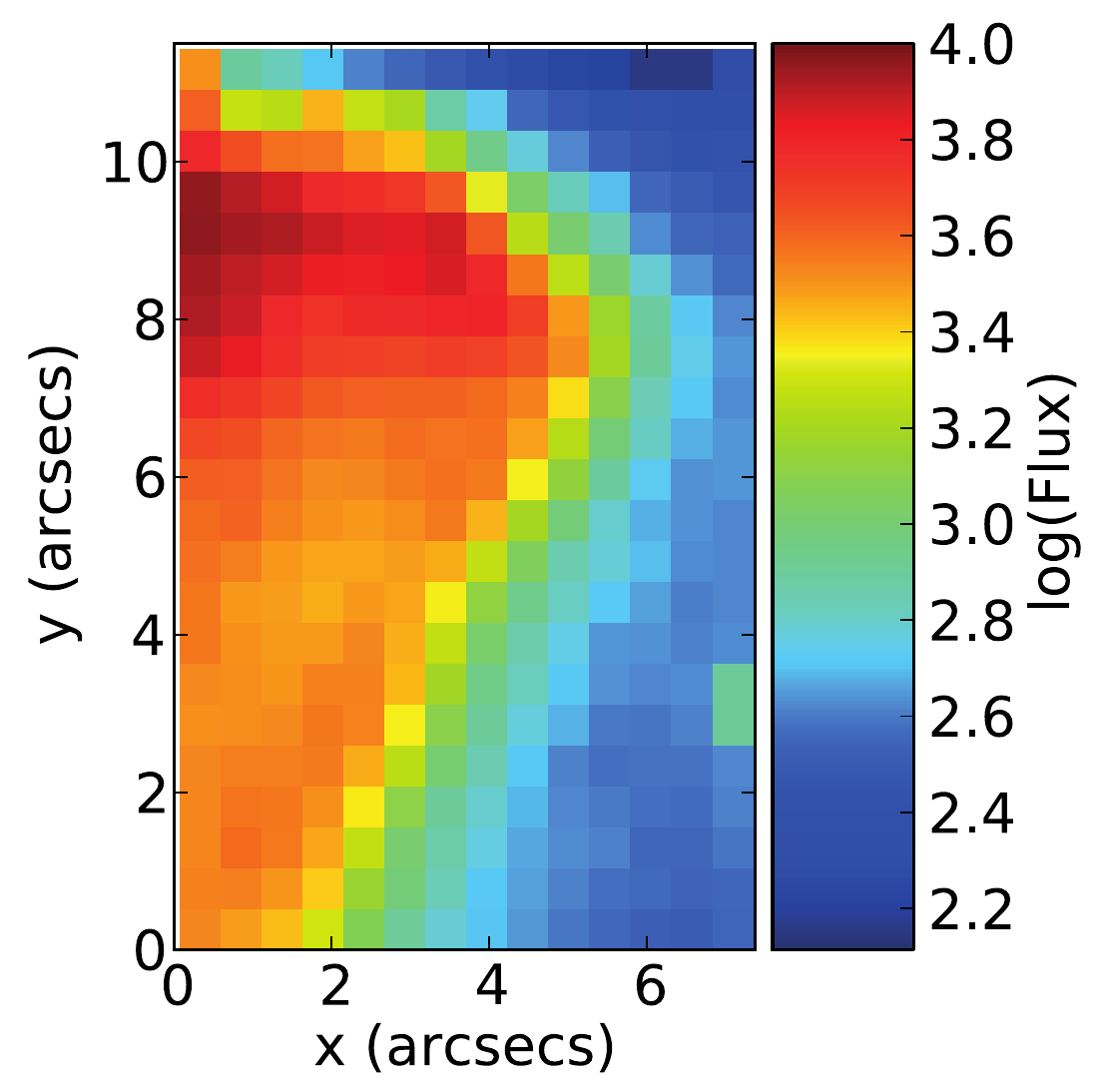}
\end{minipage}
\begin{minipage}{5cm}
\includegraphics[width=\textwidth]{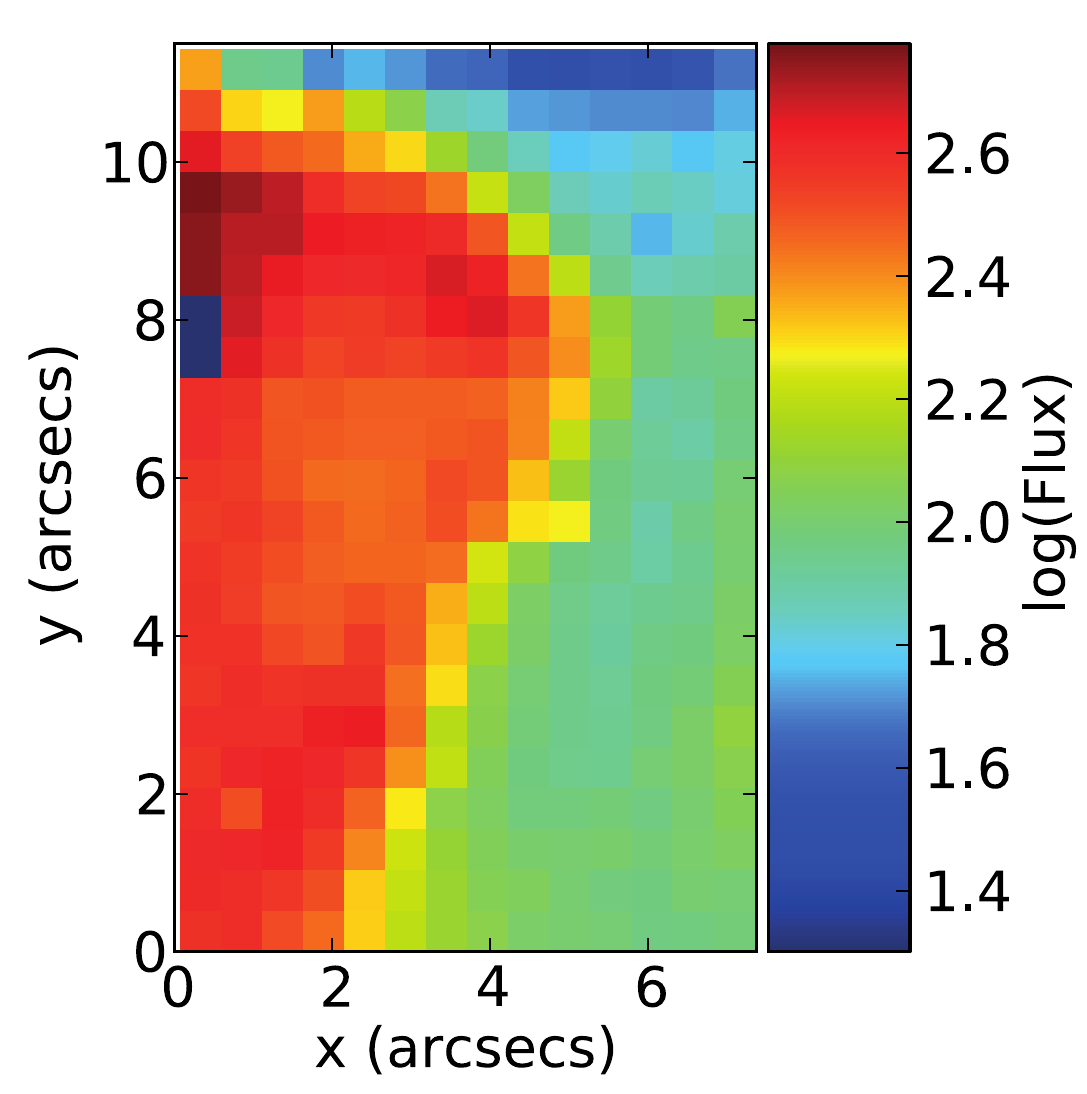}
\end{minipage}
\caption{H$\alpha$ (left), [N\two] (centre) and [S\two] (right) summed intensity maps scaled to the maximum/minimum in each map. Contours on the H$\alpha$ map represent the continuum level, and highlight the location of two stars in the field-of-view.}
\label{fig:sum_IFU2}
\end{figure*}

IFU position 2 covers the tip and part of the western edge of a pillar located to the south-west of the cluster. As for Pos1, we show maps of the integrated line flux (C1+C2) for H$\alpha$, [N\two]$\lambda$6583 and [S\two]$\lambda$6717 in Fig.~\ref{fig:sum_IFU2}. The outline of the northern and western edge of the pillar is visible in all three maps, being very clearly defined in [N\two] and [S\two]. The ratios of [N\two]/H$\alpha$ and [S\two]H$\alpha$ increase with distance from the pillar head, as discussed below in Section~\ref{sect:maps_IFU2_ratios}. Again, as for Pos1, maps of the flux, FWHM, and radial velocity for the individual identified line components in H$\alpha$ and [N\two] for Pos2 are shown in Figs.~\ref{fig:Ha_IFU2} and ~\ref{fig:NII_IFU2}, respectively. Figs.~\ref{fig:Ha_IFU2_egfits} and \ref{fig:NII_IFU2_egfits} show example line profile fits.

\subsubsection{H$\alpha$ maps} \label{sect:maps_IFU2_Ha}
Strong H$\alpha$ emission is found across the whole IFU field, but the strongest emission originates on the pillar tip. In C1, emission from the pillar is, compared to the surrounding gas, characterised by narrower line widths ($\sim$25 vs.\ 35--50~\kms) and blueshifted velocities (8--12 vs.\ 20--28~\kms), with a sharp transition at the pillar edge between the two profile shapes. The bluest C1 velocities are found on the face of the pillar, with a gradual decrease towards the edge (also seen in [N\two]). This is consistent with a radially directed outflow from the pillar surface, as would be expected from photoevaporation. Assuming that the outflow is completely radial at the bluest point and completely tangential at the pillar surface, the velocity difference would be a measure of the outflow speed. We measure a difference of $\sim$10~\kms\ (the sonic speed at $10^4$~K), which is in excellent agreement with what would be expected from photoevaporation \citep{lefloch94}.

Behind the tip of the pillar (i.e.\ downstream to the south-east), we identify a second, much broader (FWHM$\sim$55~\kms) line component whose velocity centroid is redshifted by $\sim$10--15~\kms\ compared to the corresponding narrow component (see line profile examples a and particularly b in Fig.~\ref{fig:Ha_IFU2_egfits}). Unlike in Pos1 (and in [N\two], see below), the broadest H$\alpha$ components are not found on the pillar tip or directly along its edge, but in a region interior and roughly parallel with the edge. In the area to the north-west of the pillar where only one component is identified, the line becomes comparatively broad to the broad component on the pillar, with widths up to $\sim$55~\kms. However, these line widths may simply reflect a bias in the fitting due to a faint red wing to the profile (see example Fig.~\ref{fig:Ha_IFU2_egfits}c). Unfortunately this red wing in this case is too faint to be robustly fit by our routine.

\subsubsection{[N\two] maps} \label{sect:maps_IFU2_NII}
As mentioned above, the pillar is much more clearly defined in the [N\two] flux map, with emission from the surrounding gas much fainter than that of the pillar itself. As with H$\alpha$, the C1 emission from the pillar is much narrower than that from the surroundings ($\sim$10--20 vs.\ 35--60~\kms), but the comparative blueshift of the pillar is less pronounced. All along the projected pillar edge (including the tip), we find evidence for a broad, secondary line component with FWHM$\lesssim$90~\kms\ (see example Fig.~\ref{fig:NII_IFU2_egfits}c). As alluded to above, the H$\alpha$ line profile does not exhibit these same broad components at the corresponding location: compare Fig.~\ref{fig:NII_IFU2_egfits}c to Fig.~\ref{fig:Ha_IFU2_egfits}d showing the H$\alpha$ profile from the same spaxel. Also compare Fig.~\ref{fig:Ha_IFU2_egfits}b, which shows the H$\alpha$ profile from the region with the broadest H$\alpha$ components, to Fig.~\ref{fig:NII_IFU2_egfits}b, showing the [N\two] profile from the same spaxel. Furthermore, whereas the broad component in H$\alpha$ is redshifted compared to the bright narrower component associated with the pillar, the [N\two] broad component is centred at approximately the same or bluer radial velocities compared to its corresponding narrow component. 

As with Pos1, the [S\two] line maps are very similar to the [N\two] maps but much noisier so are not shown here.

\subsubsection{Line ratios} \label{sect:maps_IFU2_ratios}
\begin{figure*}
\begin{minipage}{7.5cm}
\includegraphics[width=\textwidth]{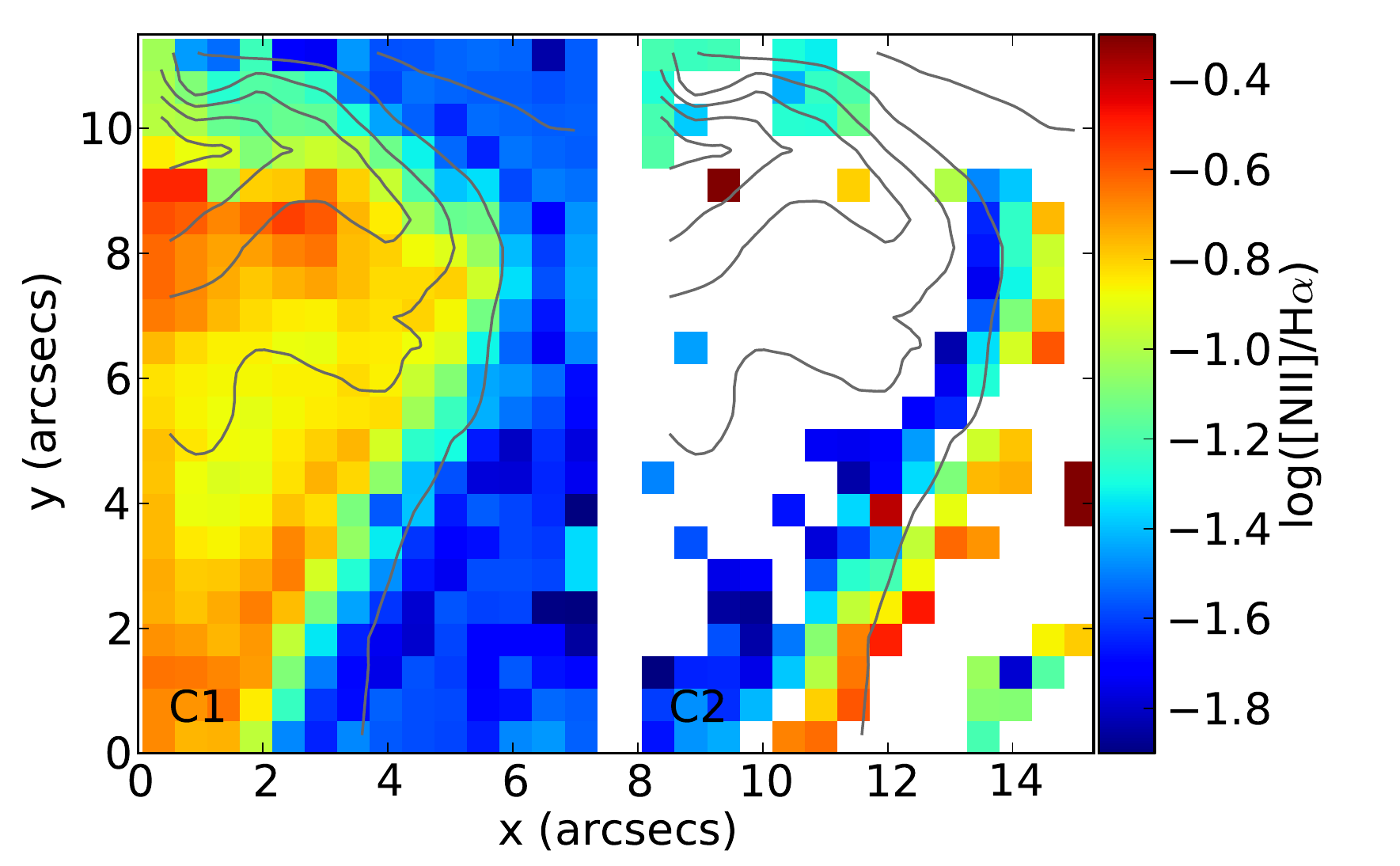}
\end{minipage}
\begin{minipage}{7.5cm}
\includegraphics[width=\textwidth]{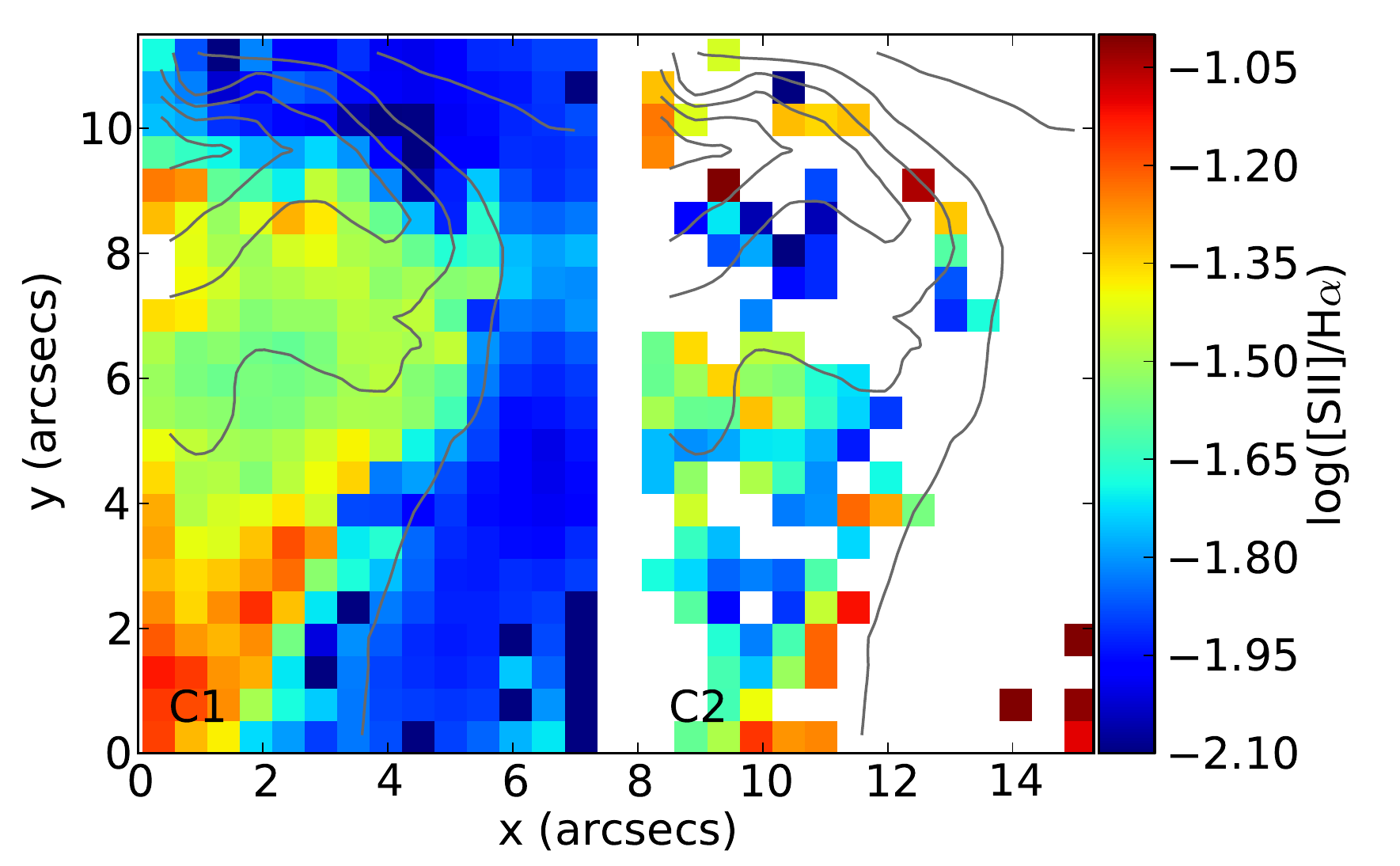}
\end{minipage}
\caption{[N\two]/H$\alpha$ and [S\two]/H$\alpha$ line ratio maps.}
\label{fig:ratios_IFU2}
\end{figure*}

We present the [N\two]/H$\alpha$ and [S\two]/H$\alpha$ line ratio maps in the two identified line components in Fig.~\ref{fig:ratios_IFU2}. Again, line ratios were only calculated where the corresponding components in both the lines were identified. In C1, the pillar exhibits much higher ratios in both [N\two]/H$\alpha$ and [S\two]/H$\alpha$ than the surrounding gas. However, the [N\two]/H$\alpha$ ratio peaks close to the tip of the pillar (at $y\sim 8$--9$''$), with a secondary peak further down on the pillar edge. [S\two]/H$\alpha$ peaks only at the secondary [N\two]/H$\alpha$ peak (downstream on the pillar edge). In C2, the highest line ratios in both indicators are found along the projected pillar edge. 

\begin{figure*}
\includegraphics[width=0.26\textwidth]{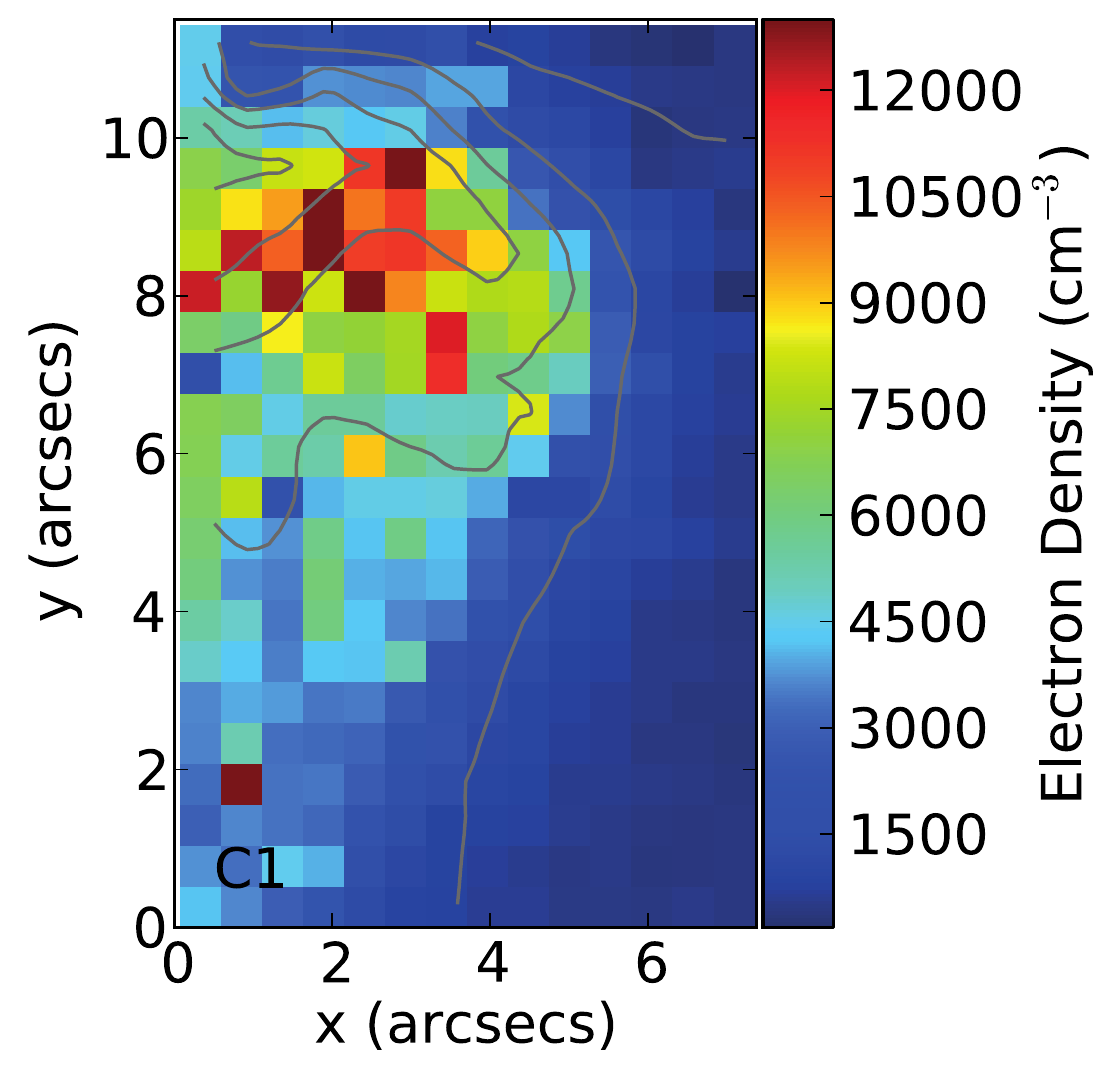}
\caption{Electron density map derived from the summed line flux ratios of the [S\two]$\lambda$$\lambda$6717,6731 doublet.}
\label{fig:IFU2_elecdens}
\end{figure*}

The electron density map derived from the summed (integrated) [S\two] line fluxes is shown in Fig.~\ref{fig:IFU2_elecdens}. The pillar tip shows up as a very high density ($>$10\,000~\cmt) region, with the density decreasing along the length of the pillar down to $\sim$2000~\cmt. The gas surrounding the pillar is at low density (at or below the low-density limit of 50--100~\cmt). Like in position 1, these high densities imply that the majority of the [S\two] emission originates from gas near to the (deeper, denser) neutral/molecular layers of the pillar and is just being ionized.

\subsubsection{PV diagram}
\begin{figure*}
\begin{minipage}{7.5cm}
\begin{overpic}[width=\textwidth]{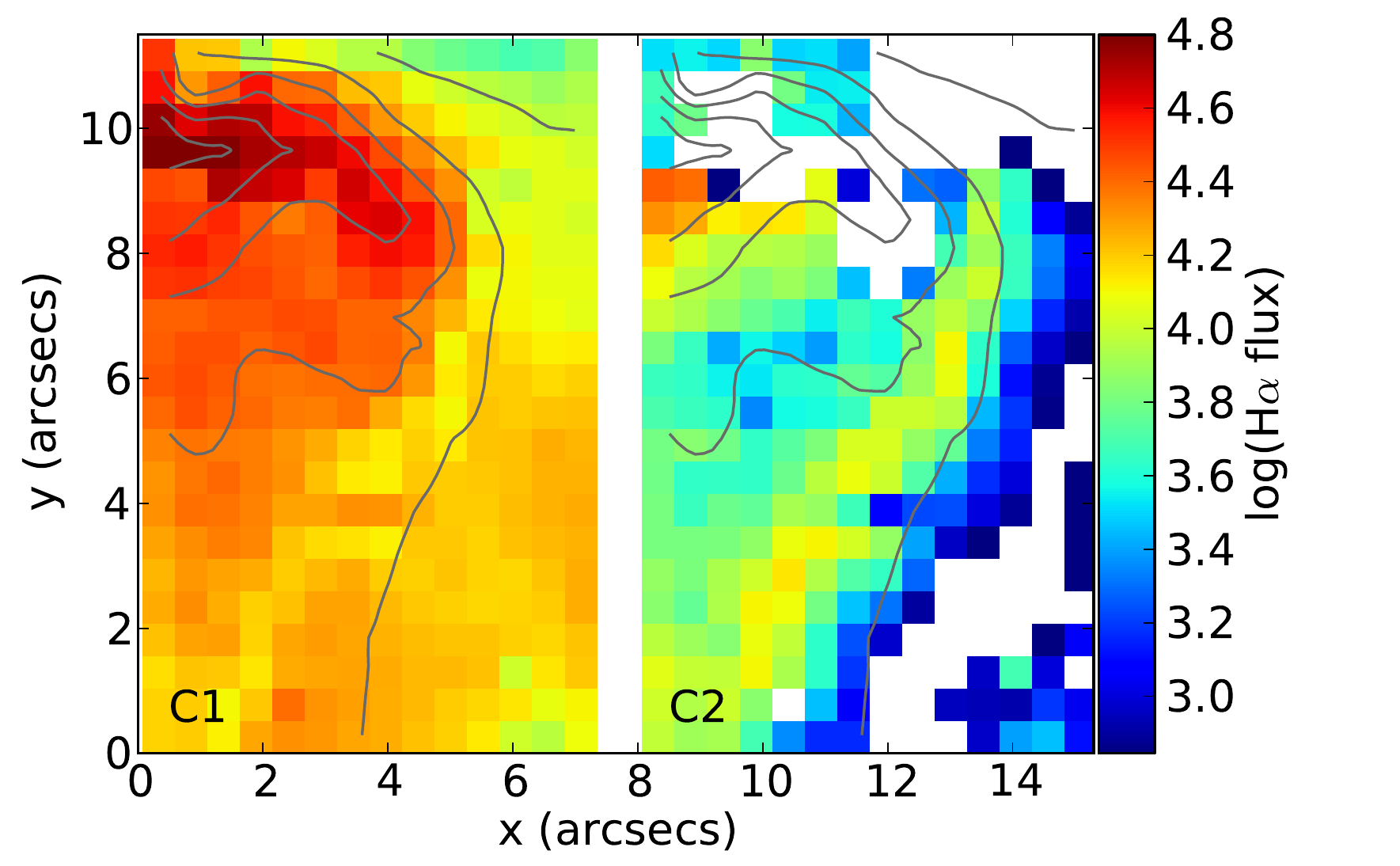}
\put(11,43){\circled{\Large{\textcolor{white}{a}}}}
\put(16,31){\circled{\Large{\textcolor{white}{b}}}}
\put(36,54){\circled{\Large{\textcolor{white}{c}}}}
\put(34,41){\circled{\Large{d}}}
\end{overpic}
\end{minipage}
\begin{minipage}{7.5cm}
\includegraphics[width=\textwidth]{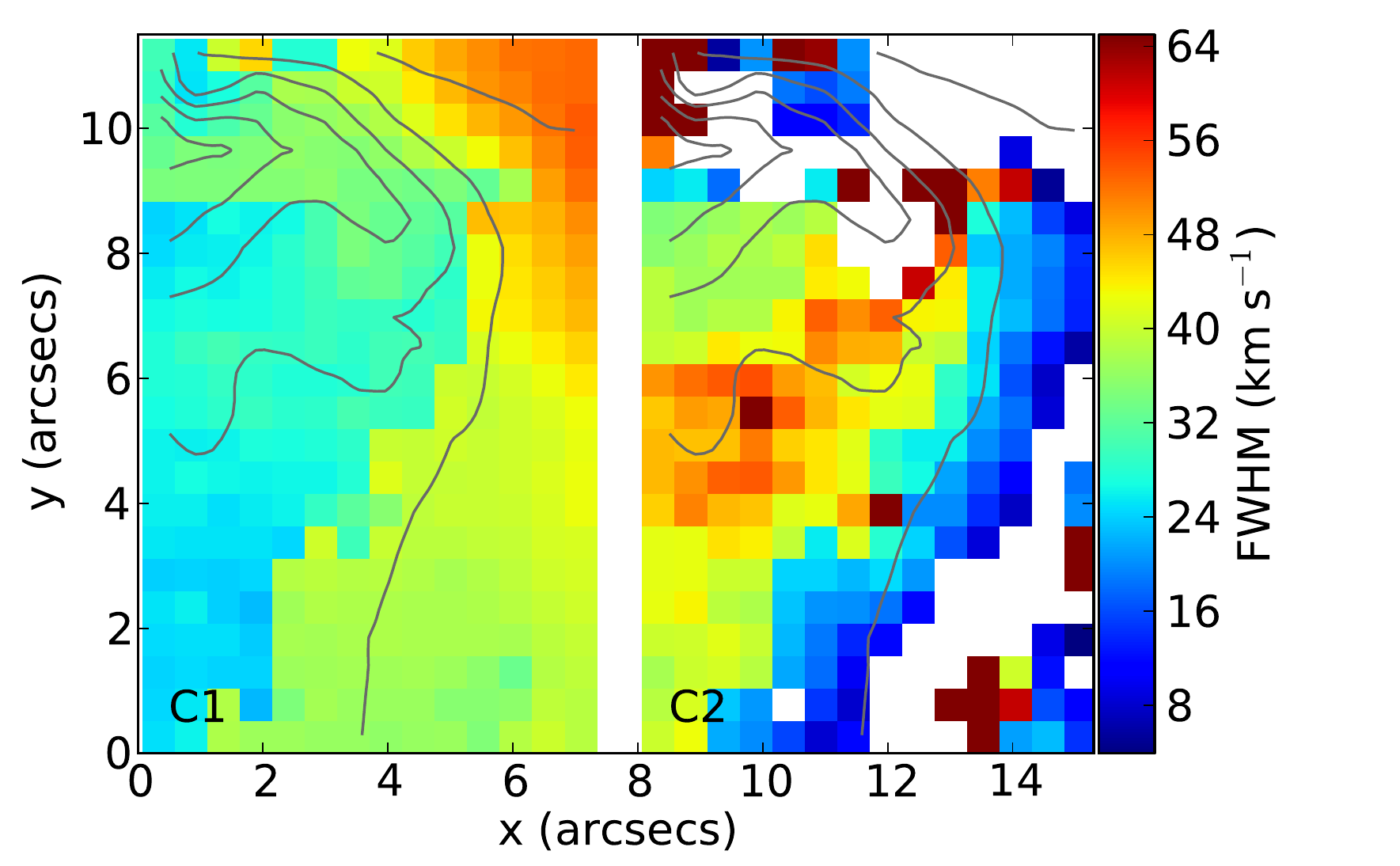}
\end{minipage}
\begin{minipage}{7.5cm}
\includegraphics[width=\textwidth]{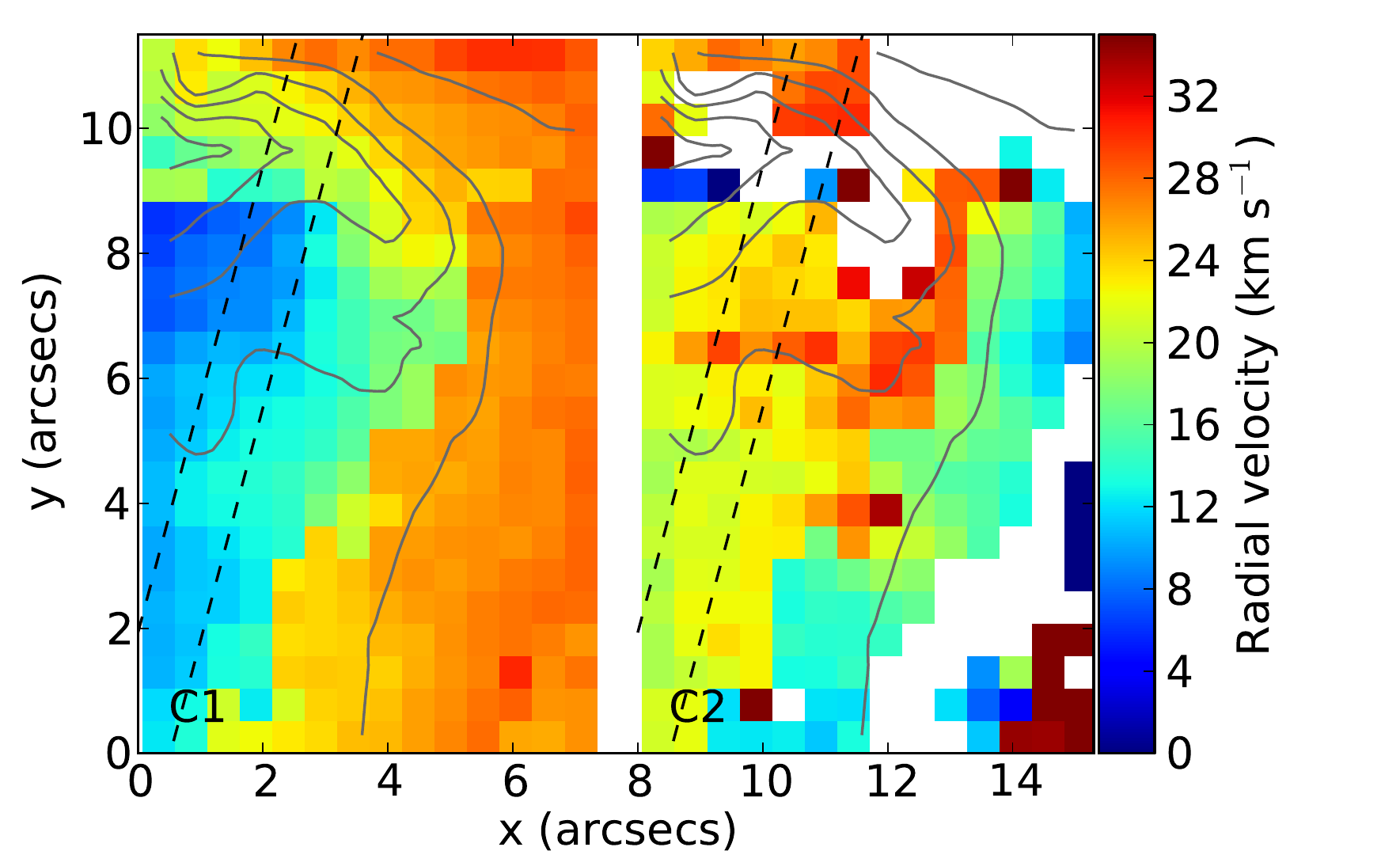}
\end{minipage}
\caption{H$\alpha$ line component maps for IFU position 2. Component 1 (C1) is shown on the left, and component 2 (C2) on the right starting at $x=8''$. The flux scale bar is in units of $10^{-18}$~erg~s$^{-1}$~cm$^{-2}$~spaxel$^{-1}$. Grey contours on this and the subsequent maps represent the total (C1+C2+C3) H$\alpha$ flux. The spaxels from which the example line profiles shown in Fig.~\ref{fig:Ha_IFU2_egfits} were extracted are labelled with the corresponding letters. The parallel dashed lines indicate the orientation and width of the pseudoslit used to extract the position-velocity information for the plot in Fig.~\ref{fig:IFU2_pv}.}
\label{fig:Ha_IFU2}
\end{figure*}

\begin{figure*}
\begin{minipage}{5.2cm}
\begin{overpic}[width=\textwidth]{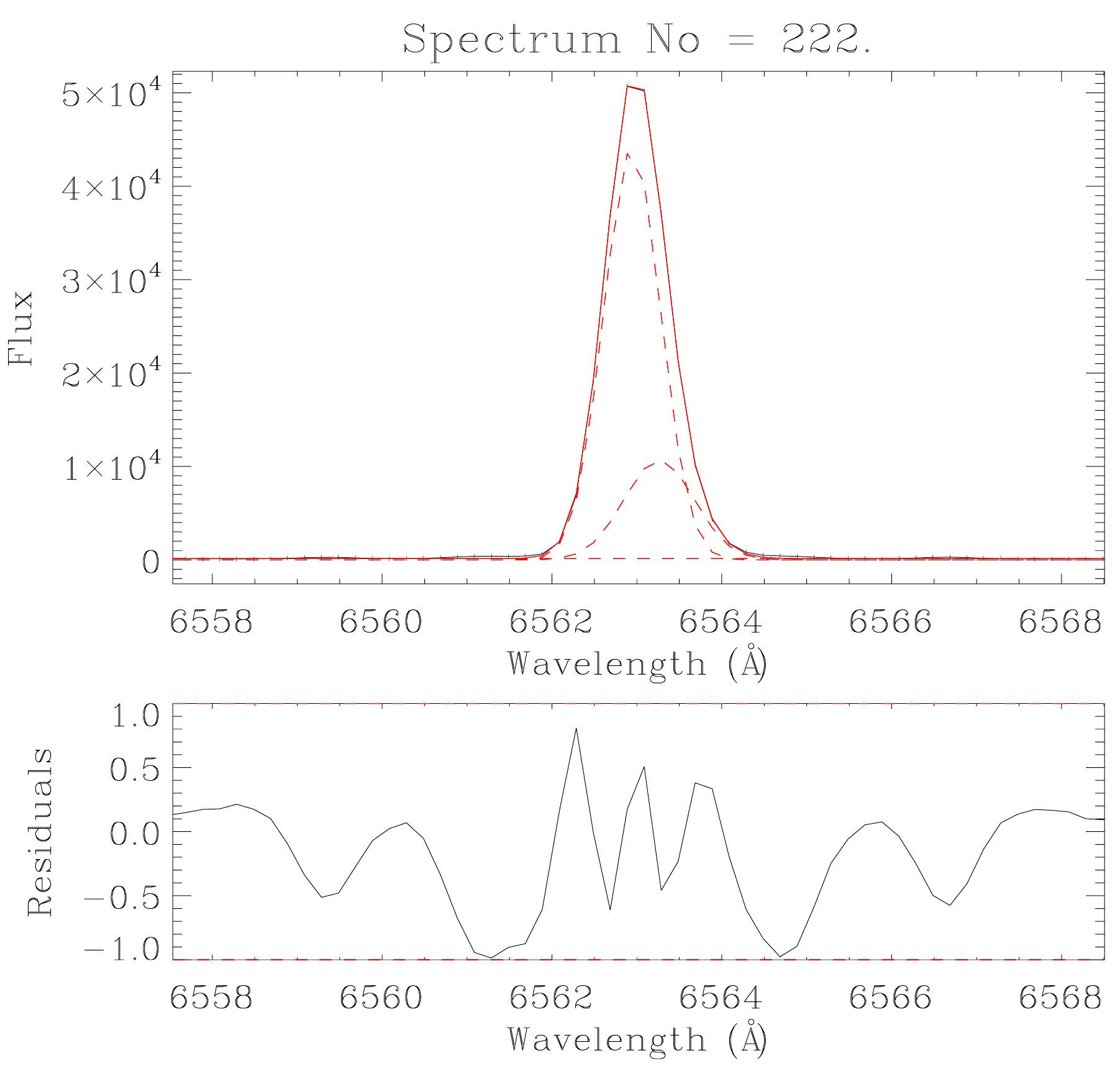}
\put(21,80){(a)}
\end{overpic}
\end{minipage}
\begin{minipage}{5cm}
\begin{overpic}[width=\textwidth]{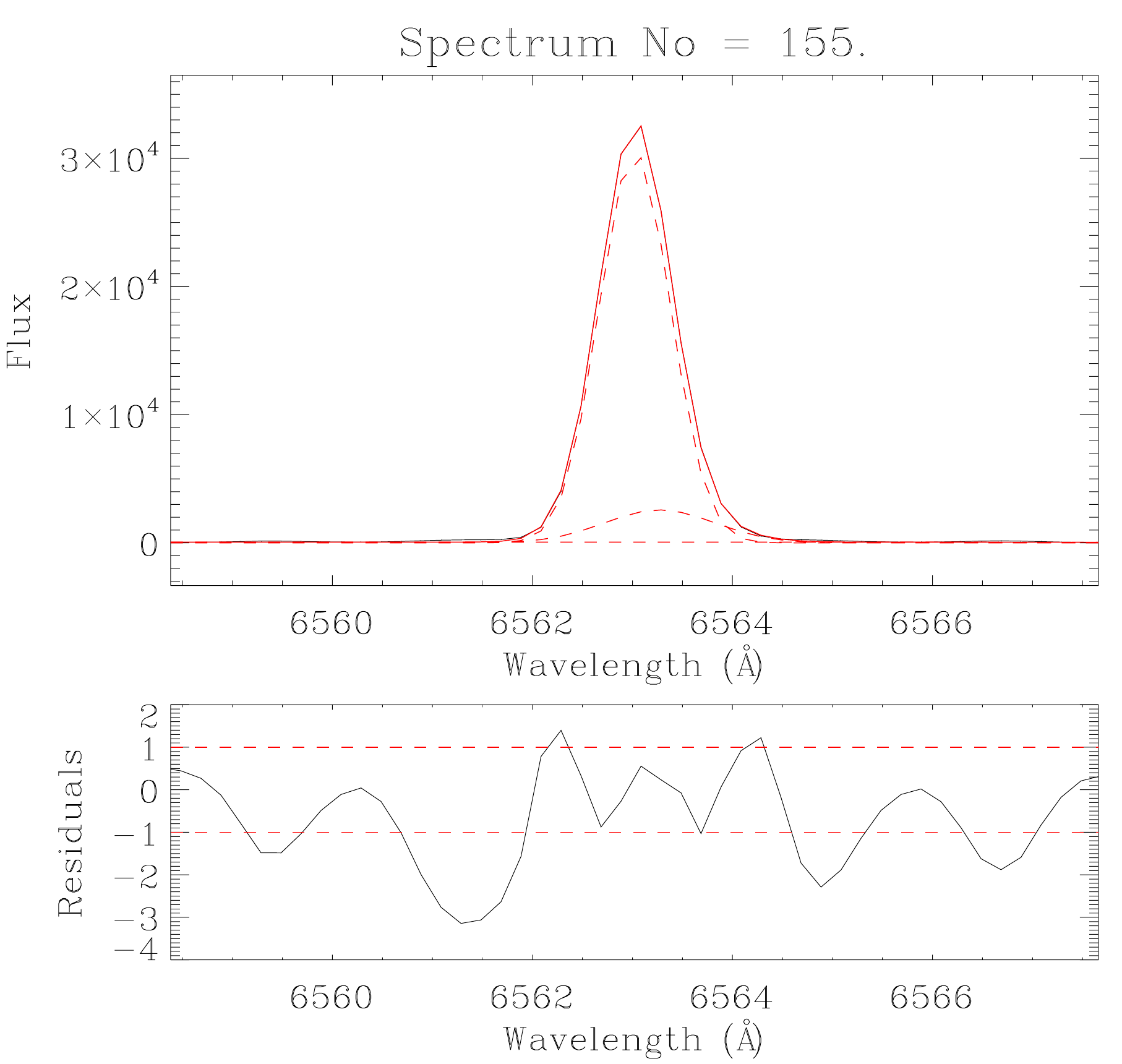}
\put(21,80){(b)}
\end{overpic}
\end{minipage}
\begin{minipage}{5cm}
\begin{overpic}[width=\textwidth]{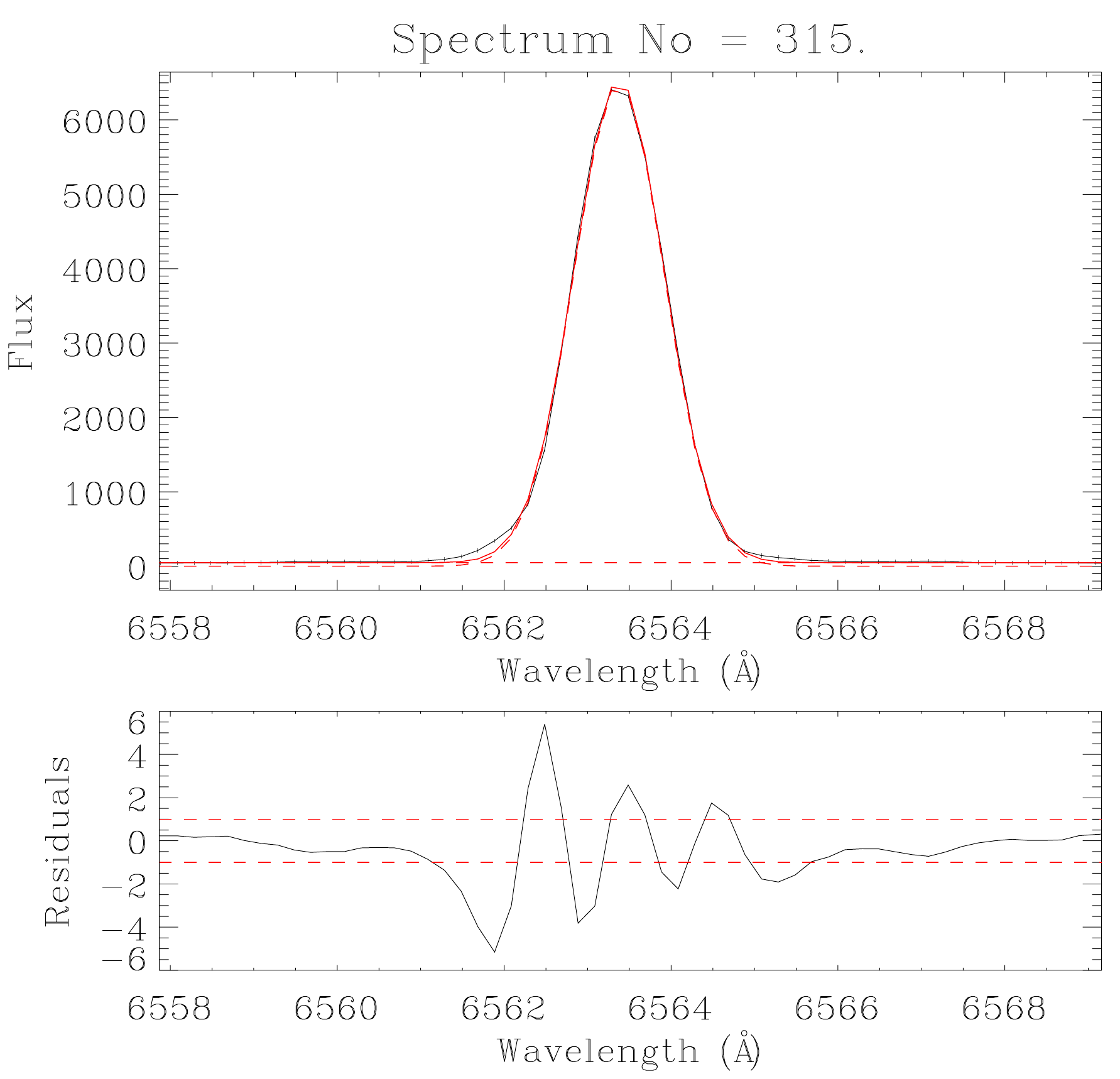}
\put(21,80){(c)}
\end{overpic}
\end{minipage}
\begin{minipage}{5cm}
\begin{overpic}[width=\textwidth]{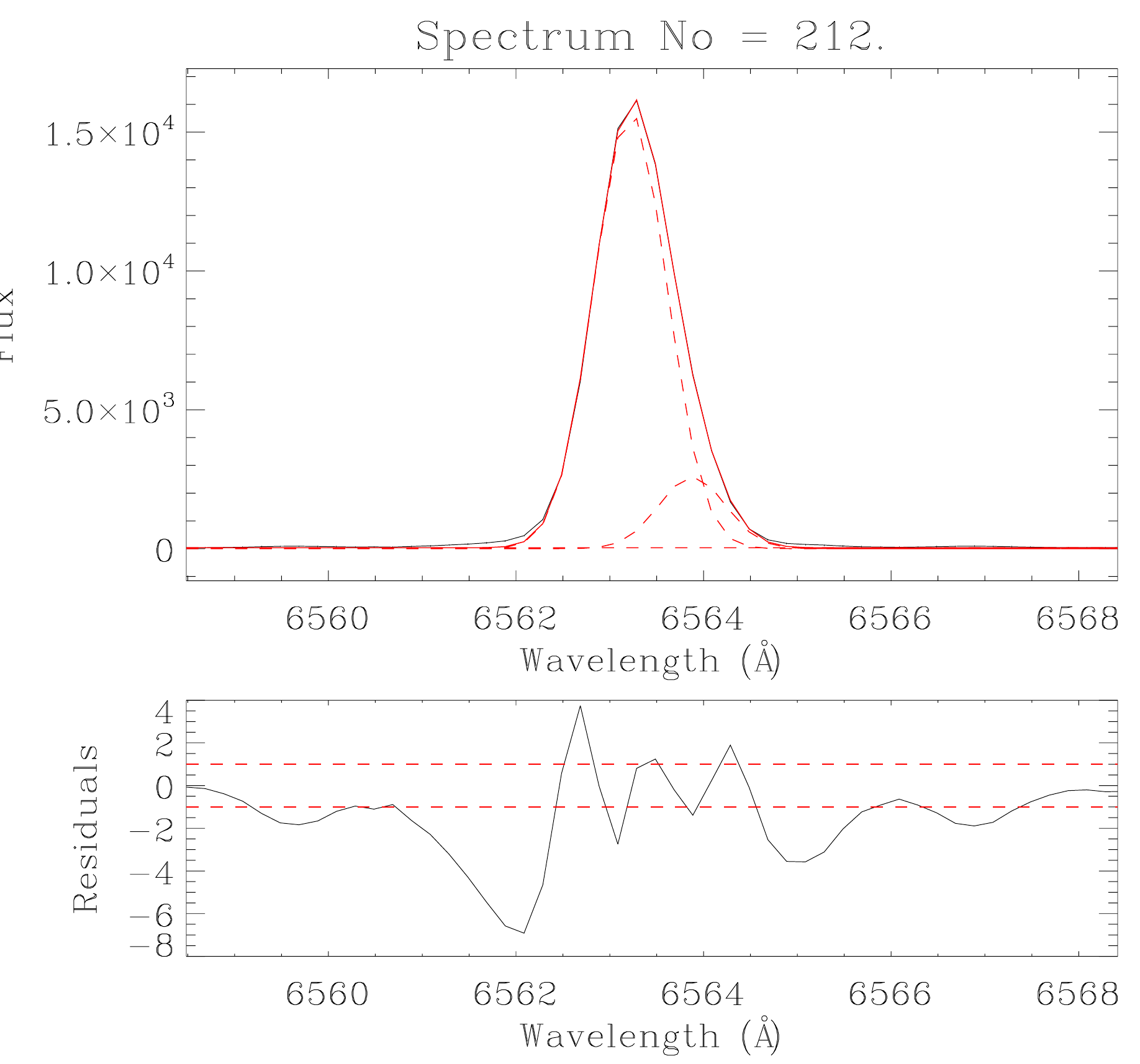}
\put(21,80){(d)}
\end{overpic}
\end{minipage}
\caption{Example H$\alpha$ line profiles extracted from the locations shown in Fig.~\ref{fig:Ha_IFU2}, chosen to represent the main types of profile shapes observed over the field. See the caption to Fig.~\ref{fig:Ha_IFU1_egfits} for further details.}
\label{fig:Ha_IFU2_egfits}
\end{figure*}

\begin{figure*}
\begin{minipage}{7.5cm}
\begin{overpic}[width=\textwidth]{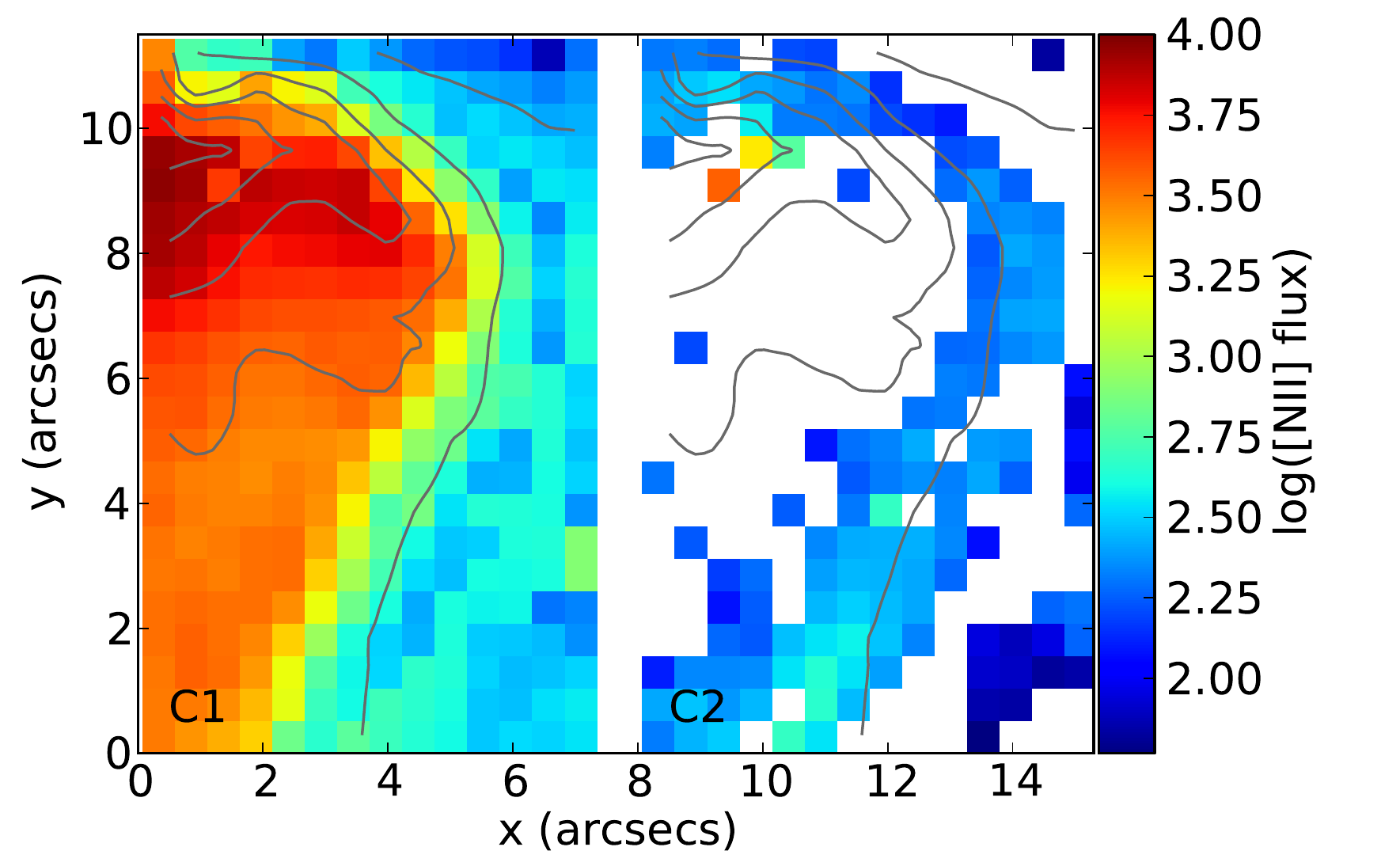}
\put(14,16){\circled{\Large{\textcolor{white}{a}}}}
\put(34,41){\circled{\Large{c}}}
\put(16,31){\circled{\Large{\textcolor{white}{b}}}}
\end{overpic}
\end{minipage}
\begin{minipage}{7.5cm}
\includegraphics[width=\textwidth]{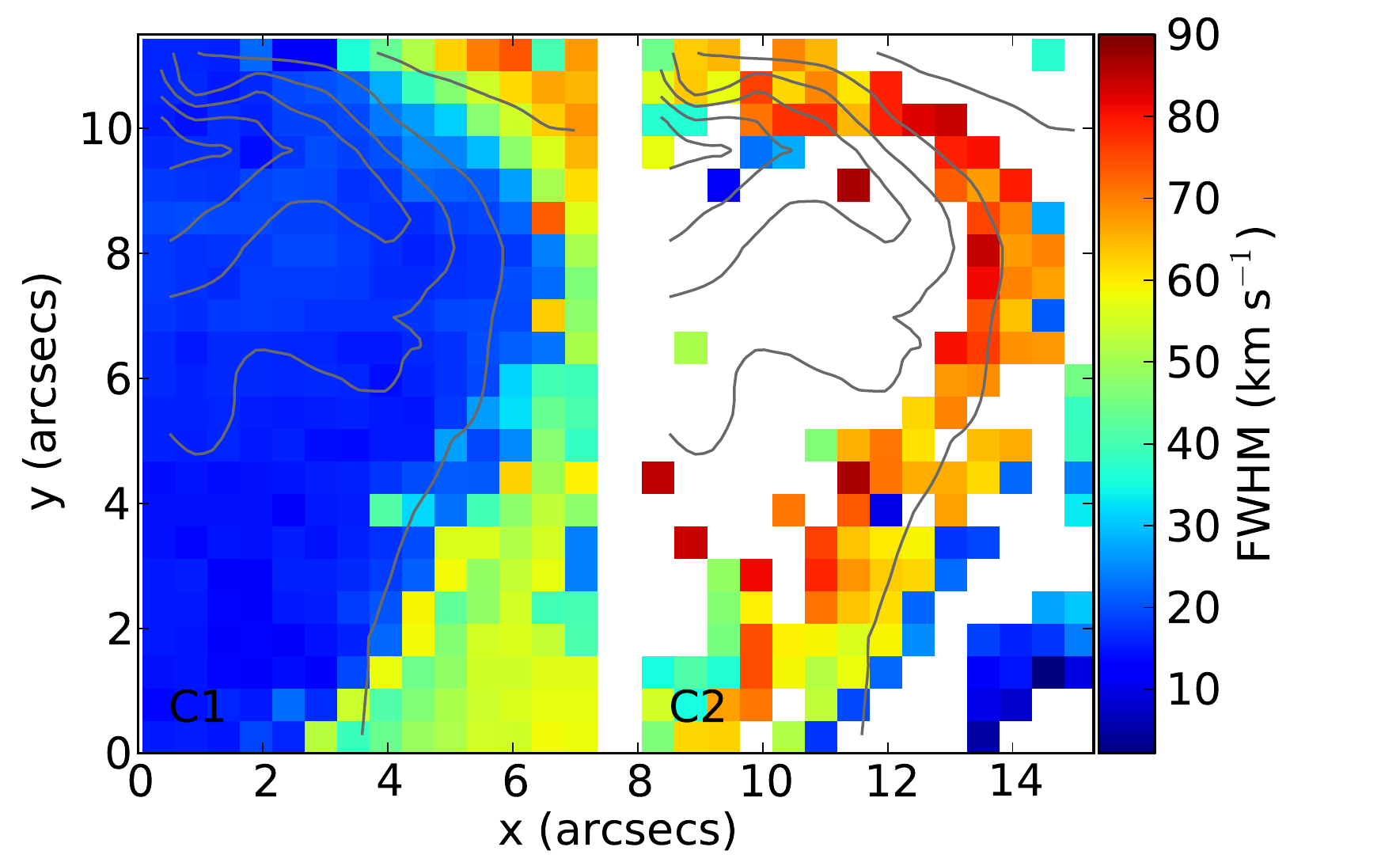}
\end{minipage}
\begin{minipage}{7.5cm}
\includegraphics[width=\textwidth]{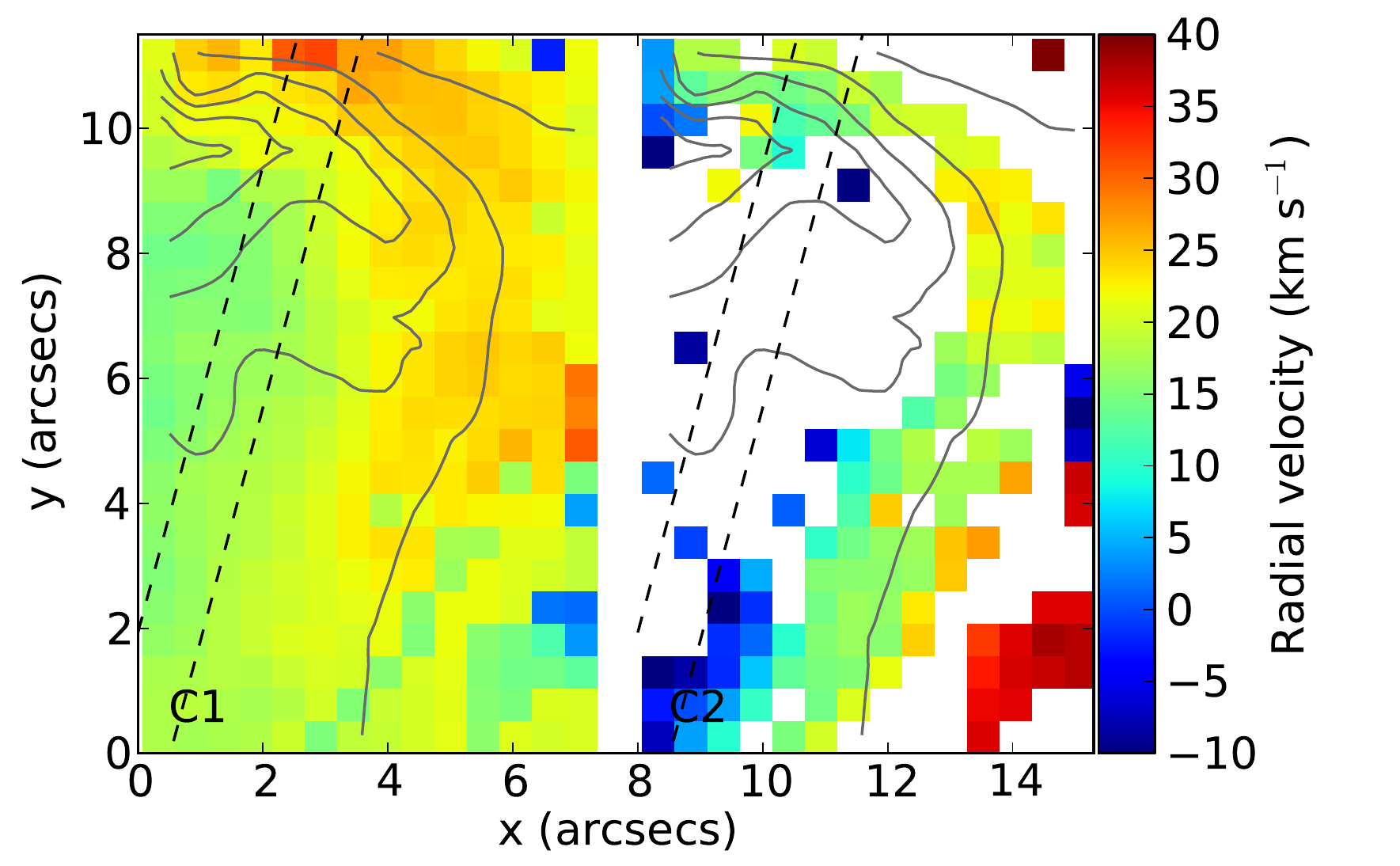}
\end{minipage}
\caption{As Fig.~\ref{fig:Ha_IFU2} but for the [N\two] line components.}
\label{fig:NII_IFU2}
\end{figure*}

\begin{figure*}
\begin{minipage}{5.2cm}
\begin{overpic}[width=0.99\textwidth]{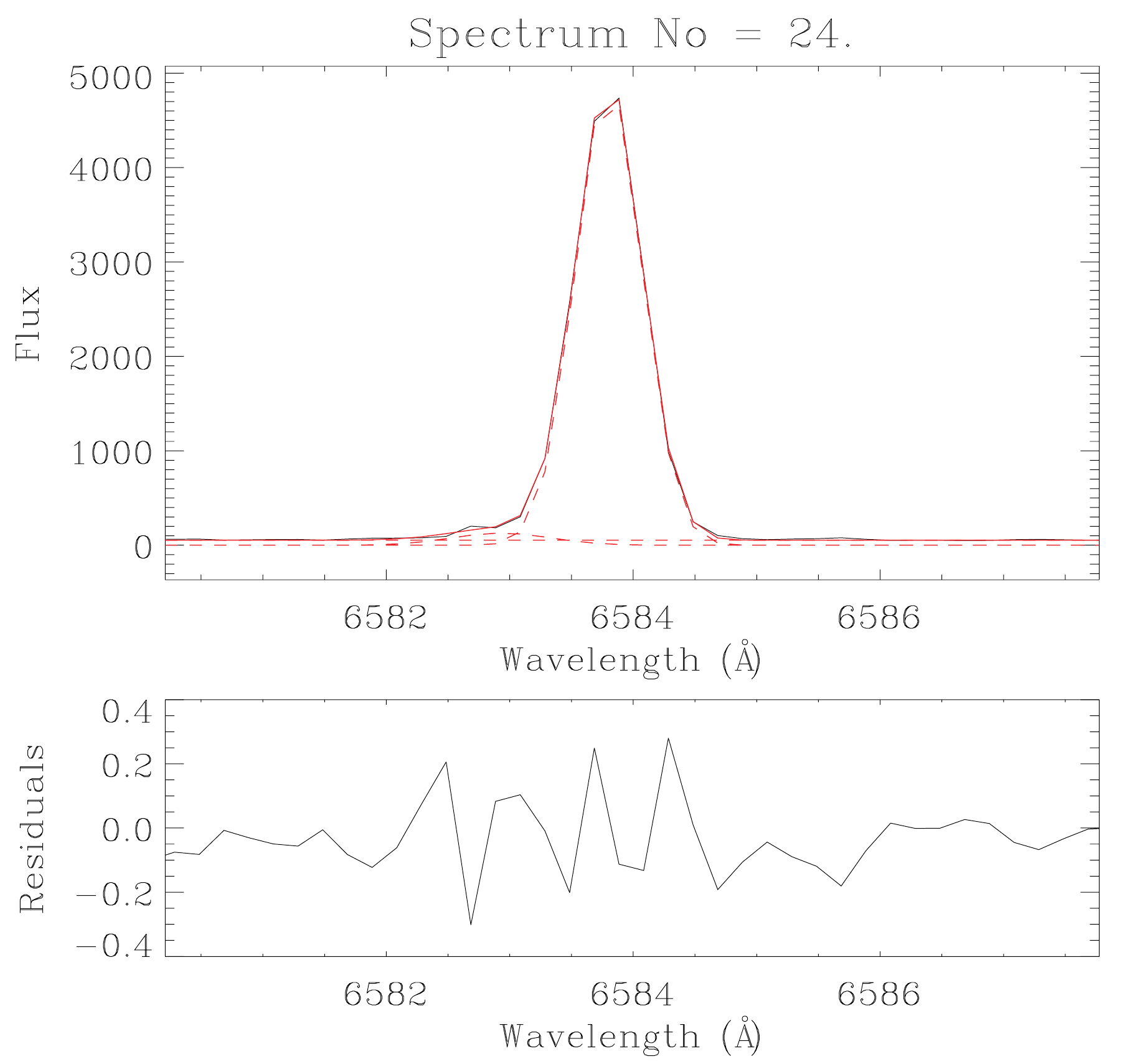}
\put(21,80){(a)}
\end{overpic}
\end{minipage}
\begin{minipage}{5cm}
\begin{overpic}[width=\textwidth]{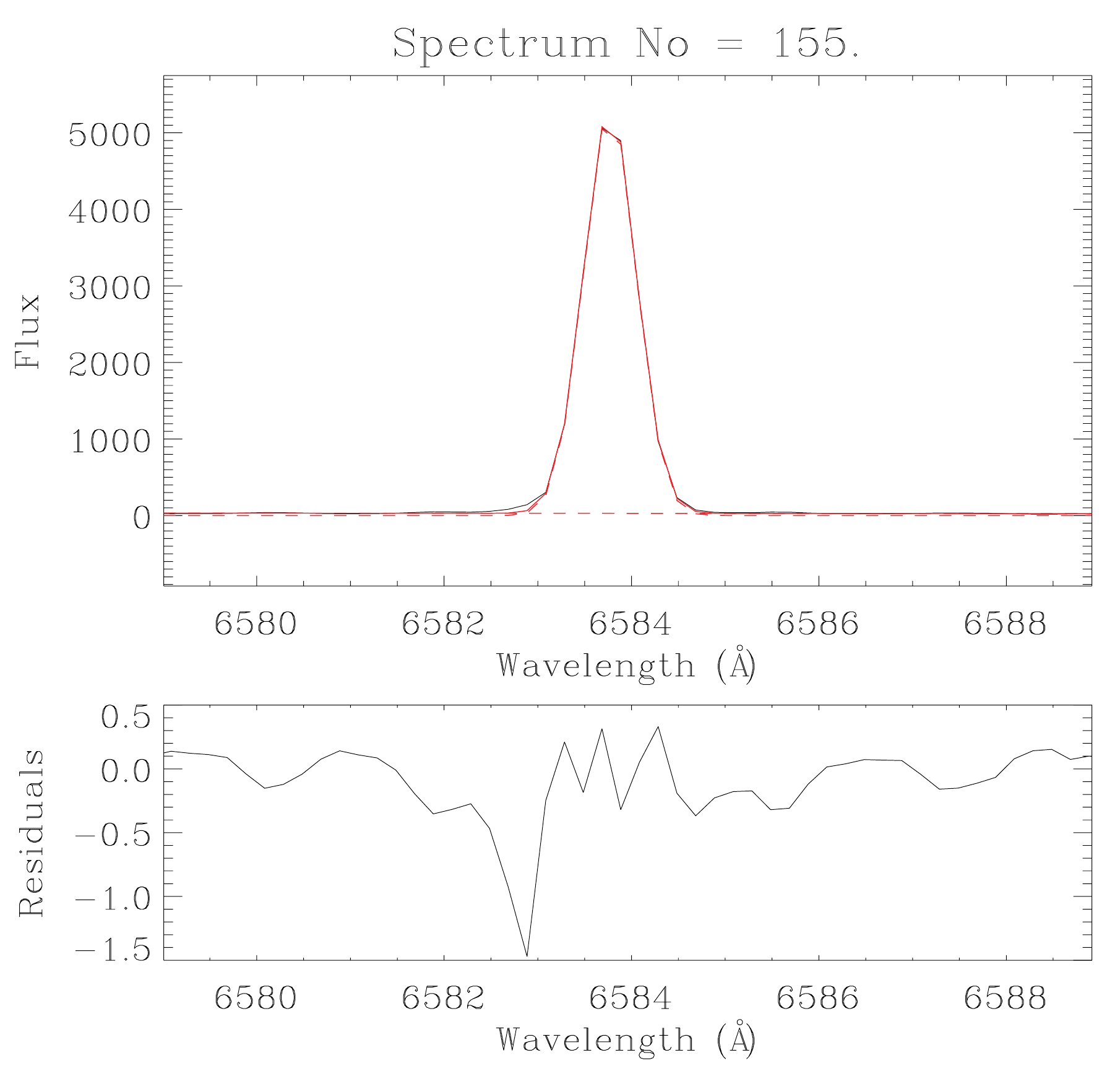}
\put(21,80){(b)}
\end{overpic}
\end{minipage}
\begin{minipage}{5cm}
\begin{overpic}[width=\textwidth]{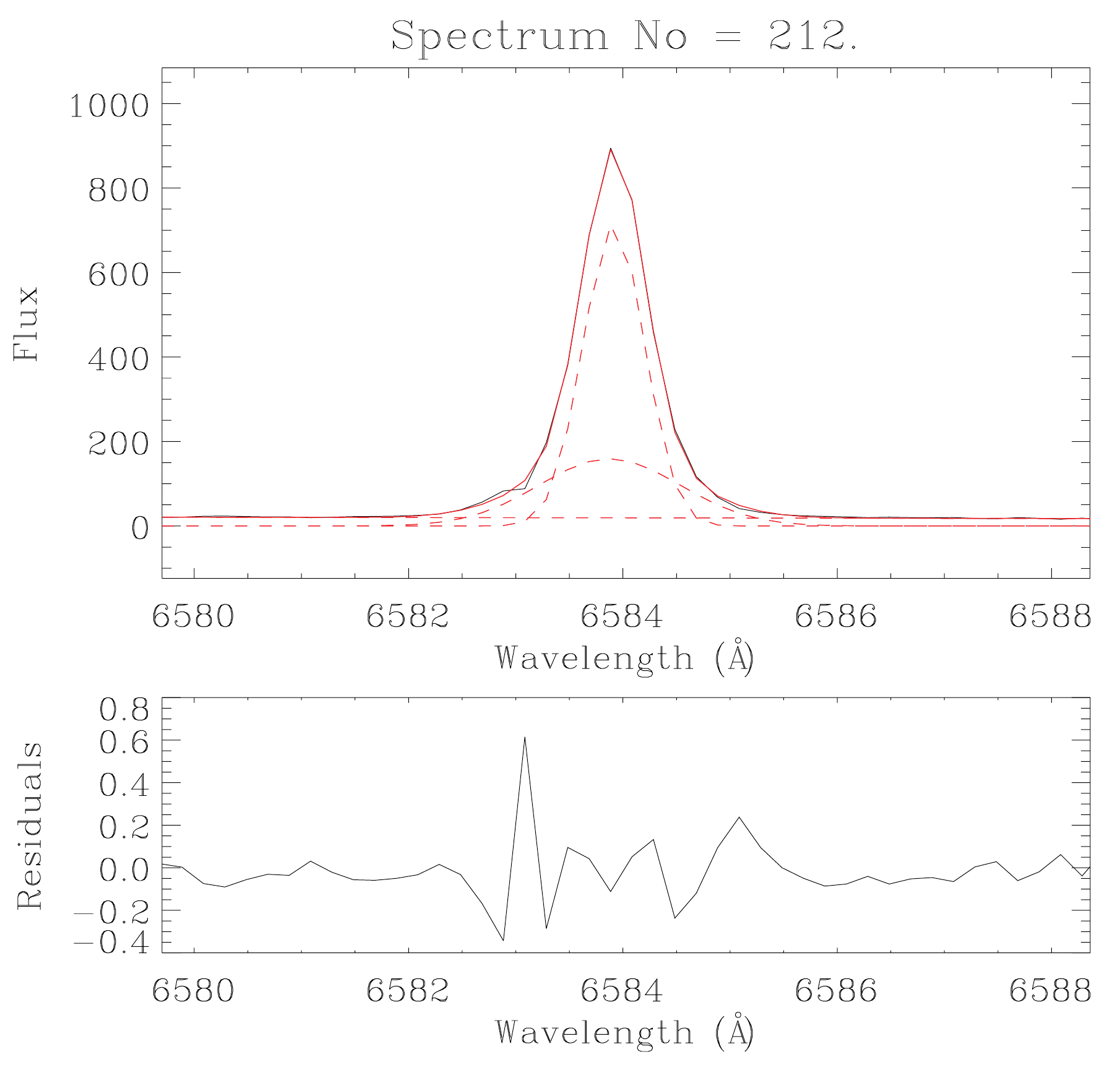}
\put(21,80){(c)}
\end{overpic}
\end{minipage}
\caption{Example [N\two] line profiles extracted from the locations shown in Fig.~\ref{fig:NII_IFU2}. See the caption to Fig.~\ref{fig:Ha_IFU1_egfits} for further details.}
\label{fig:NII_IFU2_egfits}
\end{figure*}

\begin{figure*}
\includegraphics[width=0.7\textwidth]{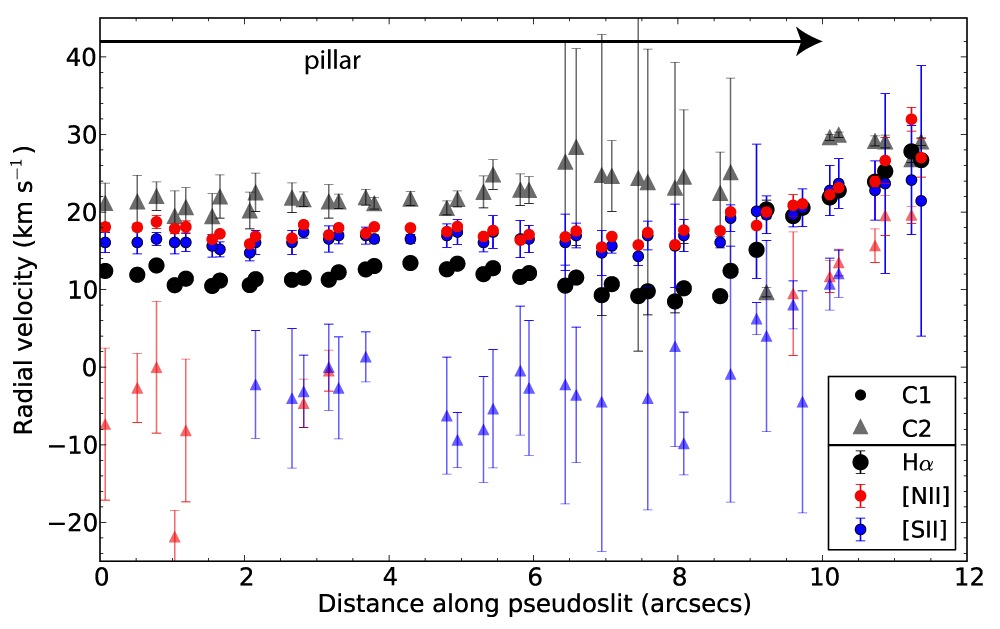}
\caption{H$\alpha$, [N\two] and [S\two] position-velocity diagram along the pseudoslit oriented along the pillar as shown in the velocity maps of Fig.~\ref{fig:Ha_IFU2} and \ref{fig:NII_IFU2}. Circles represent C1 and triangles C2, and the 3 colours represent Halpha (black), [N\two] (red) and [S\two] (blue). Radial velocities are in heliocentric units. Distances on the $x$-axis are measured from the bottom-left of the IFU field, increasing towards the tip of the pillar (at $\sim$10--11$''$). Error bars on each point reflect the uncertainties given in the \textsc{pan} fits, and for C1 are generally smaller than the symbol size.}
\label{fig:IFU2_pv}
\end{figure*}

In order to examine the radial velocity distribution along the pillar in this position in more detail, we have defined a ``pseudoslit'' oriented along the long axis of the pillar from which we have extracted the velocity information for H$\alpha$, [N\two] and [S\two]. The location and width of the pseudoslit is shown on the radial velocity maps in both Figs.~\ref{fig:Ha_IFU2} and \ref{fig:NII_IFU2}, and the extracted position-velocity data are plotted in Fig.~\ref{fig:IFU2_pv}. For all three lines, the C1 velocities remain remarkably constant (within 4~\kms) along the pillar, however there is clearly a consistent offset between H$\alpha$ and [N\two] (and [S\two]) of $\sim$5--8~\kms. Beyond $\sim$9$''$ (the pillar tip and beyond), the C1 velocities of H$\alpha$, [N\two] and [S\two] converge and all three increase up to $\sim$27~\kms. On the pillar, H$\alpha$ C2 (black triangles) is redshifted from C1 by $\sim$10--15~\kms, whereas (where detected) the [N\two] and [S\two] second components are redshifted by $\sim$20~\kms.

\begin{figure*}
\includegraphics[width=0.7\textwidth]{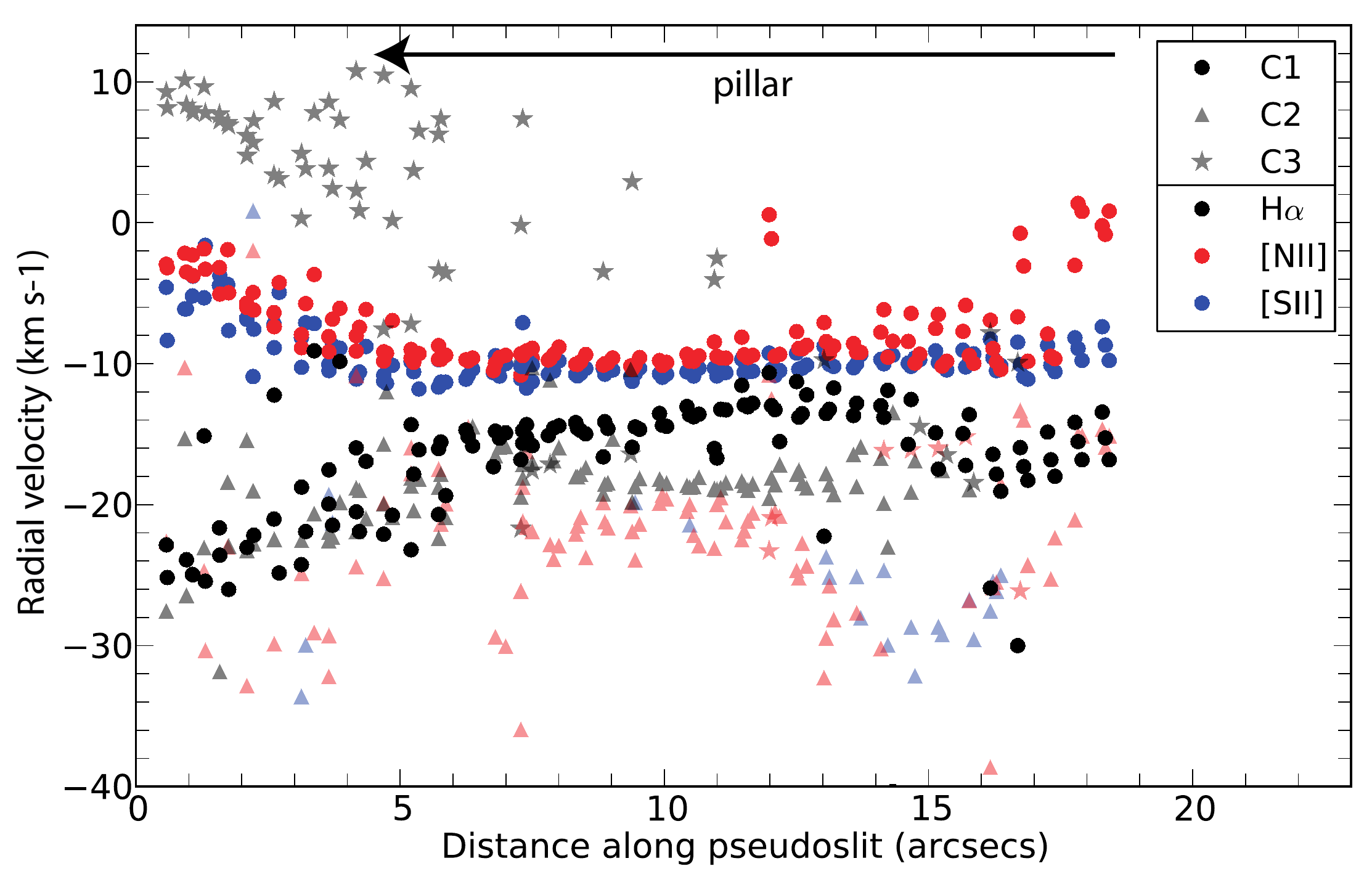}
\caption{H$\alpha$, [N\two] and [S\two] position-velocity diagram along a pseudoslit oriented along the pillar in NGC 6357 \citep[derived from re-analysed data originally presented by][]{westm10b}. Circles represent C1, triangles C2, and stars C3, and the 3 colours represent Halpha (black), [N\two] (red) and [S\two] (blue). Distances on the $x$-axis are measured from the bottom of the IFU field \citep[see][]{westm10b}; the pillar begins with its tip at $\sim$$5''$.}
\label{fig:N6357_pv}
\end{figure*}

A similar $\sim$5--10~\kms\ difference between the narrow components of H$\alpha$ and [N\two] (and [S\two]) is found on the pillar in Pos1 (PV diagram not shown). In addition, further analysis of the data presented in \citet{westm10b} of a similar pillar in NGC~6357 also shows a 5--8~\kms\ offset between H$\alpha$ and [N\two] (+ [S\two]); see Fig.~\ref{fig:N6357_pv}.


We can assume that the bright C1 component (comprising the vast majority of the line emission) originates in ionized surface layers that are accelerating radially outwards due to photoevaporation \citep[the outflow must accelerate due to the density gradient; see e.g.][]{hester96}. The outflow speed will be of order the sound-speed in the ionized gas, $\sim$10~\kms. (C2 then traces TMLs with turbulence $>>$sound speed.) On the face of the pillar, blueshifted velocities would therefore originate from material further out from the pillar core. However, along sight-lines towards the pillar edge, the photoionized outflow would primarily be moving in the tangential plane, and therefore less visible. Assuming that the pillar is a temperature-stratified structure, material further out would also be cooler and less dense (the density map supports this -- the electron density of the pillar is $\sim$10\,000~\cmt\ and falls rapidly to $<$100~\cmt\ in the surrounding gas).

That H$\alpha$ is blueshifted would suggest that it originates further out in the flow from [N\two] and [S\two]. H$\alpha$, being a recombination line, can be produced wherever there are ionized H atoms and free electrons for recombination, i.e.\ in a large range of densities and temperatures. However, the flux of H$\alpha$ scales in proportion to $n_{\rm e}^2$, so its luminosity should be weighted to the higher density (deeper) layers. [N\two] and [S\two] would then have to originate from even deeper layers. Since the densities we measure (from the [S\two] lines) are high, collisional de-excitation (quenching) may have an effect on the forbidden lines. At $10^4$~K, the critical density of [N\two]$\lambda$6583 is $\sim$80\,000~\cmt, whereas [S\two]$\lambda$6717 and [S\two]$\lambda$6731 are only 1400 and 3600~\cmt, respectively. The [S\two]$\lambda$6731 flux relative to H$\alpha$ at a density of 10\,000~\cmt\ would be approximately half of what it would be at 100~\cmt, all else being equal, while the [N\two]$\lambda$6583 flux would be about 10\% less. However, that both these ions show the same velocity structure rules out any major effect from this on the velocities. Besides, quenching of emission from the deeper layers would bias the emission towards gas that is further out, but this is the opposite of what we find.


One final, if unlikely, explanation is chemical differences between the pillar gas that is being photoevaporated and the surrounding medium. If the pillar is H-deficient, then the H$\alpha$ emission from the deeper layers would be weak. If the H-deficient flow then encounters more H-rich material, then as they mix the total emission would be biassed towards the outer, more mixed layers. There is no reason why, however, the pillar gas should be H-poor.

Thus, we are unable to explain the differences in radial velocities between H$\alpha$, [N\two] and [S\two]. Further investigation of multiple transitions probing the different temperature and density layers at high resolution is required.

\subsubsection{Summary}
In H$\alpha$ we find a broad (FWHM$\sim$55~\kms) component along the interior of the pillar edge, whose velocity centroid is redshifted with respect to the corresponding narrow component by $\sim$10--15~\kms. In [N\two], however, we find a much broader (FWHM=70--90~\kms) component located this time directly on the pillar edge, including the tip. Its centroid radial velocities are roughly consistent with that of the narrow component on the pillar itself, suggesting that these [N\two] broad components are associated with the pillar.

The bluest H$\alpha$ and [N\two] C1 velocities are found on the face of the pillar, with a gradual decrease towards the edge. This is consistent with a radially directed outflow from the pillar surface, as would be expected from photoevaporation. The velocity difference between the radial (bluest) point in the centre of the pillar face and the tangential point at the pillar surface is consistent with an outflow speed of $\sim$10~\kms\ (the sonic speed at $10^4$~K), which is in excellent agreement with what would be expected from photoevaporation.

On the pillar, [N\two]/H$\alpha$ peaks both at the pillar tip and further down near the edge, whereas [S\two]/H$\alpha$ only exhibits the latter peak. The pillar tip shows up as a very high density ($>$10\,000~\cmt) region, with the density decreasing along the length of the pillar down to $\sim$2000~\cmt. The gas surrounding the pillar is at low density (at or below the low-density limit of 50--100~\cmt). Like in position 1, these high densities imply that the majority of the [S\two] emission originates from gas near to the (deeper, denser) neutral/molecular layers of the pillar that is just being ionized.

That the radial velocities of the gas surrounding the pillar are redshifted by 20--25~\kms\ compared to the pillar, and this component is not seen at the location of the pillar, suggests that it is in the background. The electron density of this material is also at or below the low density limit ($\sim$100~\cmt), but, unlike Pos1, the [N\two]/H$\alpha$ and [S\two]/H$\alpha$ ratios are significantly lower than that of the pillar. It is unlikely, therefore, that this component equates to the material in the far background seen in the eastern edge of IFU position 1. Its radial velocities place it in the middle-distance; indeed Fig.~\ref{fig:finder} shows that there is a region of diffuse emission surrounding this pillar, which is likely to be the material we are tracing here.

As shown in the p-v diagram extracted from the pseudoslit, we find a consistent offset in radial velocity between the narrow (brighter) components of H$\alpha$ and [N\two] (+ [S\two]) of $\sim$5--8~\kms. Interestingly, this is seen in both the pillars observed here and in further analysis of the data presented in \citet{westm10b} of a similar pillar in NGC~6357. We are unable to find a satisfactory explanation for these differences in radial velocities, although the answer most likely lies in the ionization, temperature and density stratification of the pillar surface layers and the location within these that the emission lines originate.

\section{Discussion and Conclusions}
The extremely high S/N (up to $\sim$20\,000 in H$\alpha$) and spectral resolution of these data have allowed us to quantify the emission line profile shapes to a high degree of accuracy. We fitted multiple Gaussian profile models to the H$\alpha$, [N\two]$\lambda$6583 and [S\two]$\lambda$$\lambda$6717,6731 emission lines, and found a variety of narrow and broad emission components from the different regions observed.

\subsection{Narrow and broad component emission}
In all three pillars \citepalias[that in NGC~6357][and the two here in NGC~3603]{westm10b}, we find enhanced narrow line component line ratios on the pillar edges and a density peak at the pillar tips. The two pillars in NGC~3603 presented here exhibit very high ionized gas densities indeed, $>$10\,000~\cmt. In the Pos1 pillar, our observations were of a sufficiently high S/N ratio to decompose the [S\two] lines into two components, and we found that the high densities are only found in the narrow component. This implies that the narrow component originates from gas near to the (deeper, denser) neutral/molecular layers of the pillar (that is just being ionized) whereas the broad component originates in material further out. In \citetalias{westm10b} we explained these higher line ratios and densities as the effect of an ionization front passing into the pillar. However, some contribution to these observables may also come from a shock generated by the rocket effect of a photoevaporation flow that is propagating back into the pillar.

On the NGC~3603 Pos2 pillar we found the bluest H$\alpha$ and [N\two] C1 velocities on the face of the pillar, with a gradual decrease towards the edge. The existence of this gradient, and its amplitude, is consistent with a radially directed outflow from the pillar surface driven by photoevaporation. Photoevaporation streamers can also be seen in the H$\alpha$ images of the two pillars shown in Fig.~\ref{fig:finder}, as was also found in the famous images of the pillars in M16 \citep{hester96}.

The narrow C1 component, therefore, appears to trace material fairly deep within the ionized layers of the pillar surface, and exhibits the kinematic signatures of an outflow driven by photoevaporation. It comprises the vast majority of the total line emission in both H$\alpha$ and [N\two], which reflects the density dependence of the emission line strength of both lines (the critical density of [N\two]$\lambda$6583 is $\sim$80\,000~\cmt\ so it does not suffer from quenching at these densities). The line ratios suggest excitation due to a combination of an ionization and shock front propagating into the pillar.

In all three pillars we also find a broad emission component. We only observed the tip in the NGC~3603 Pos1 pillar, but for the Pos2 pillar and in NGC 6357 we find that the broad components are found close to or on the pillar edge.
In \citetalias{westm10b} we associated this broad component with turbulent mixing layers (TMLs) on the pillar surface driven by the effect of the hot, fast wind from the star cluster. The data presented here supports this conclusion. For these conditions to persist (i.e.\ in a quasi-steady-state), the photoevaporation flow and the wind must come into pressure equilibrium, and since the pressure in the shocked wind is very high, the layer of turbulent photoevaporated gas (the TML) must be quite thin. In a sense, the TML may act to ``protect'' the pillar from the full force of the wind.

\begin{figure*}
\includegraphics[width=\textwidth]{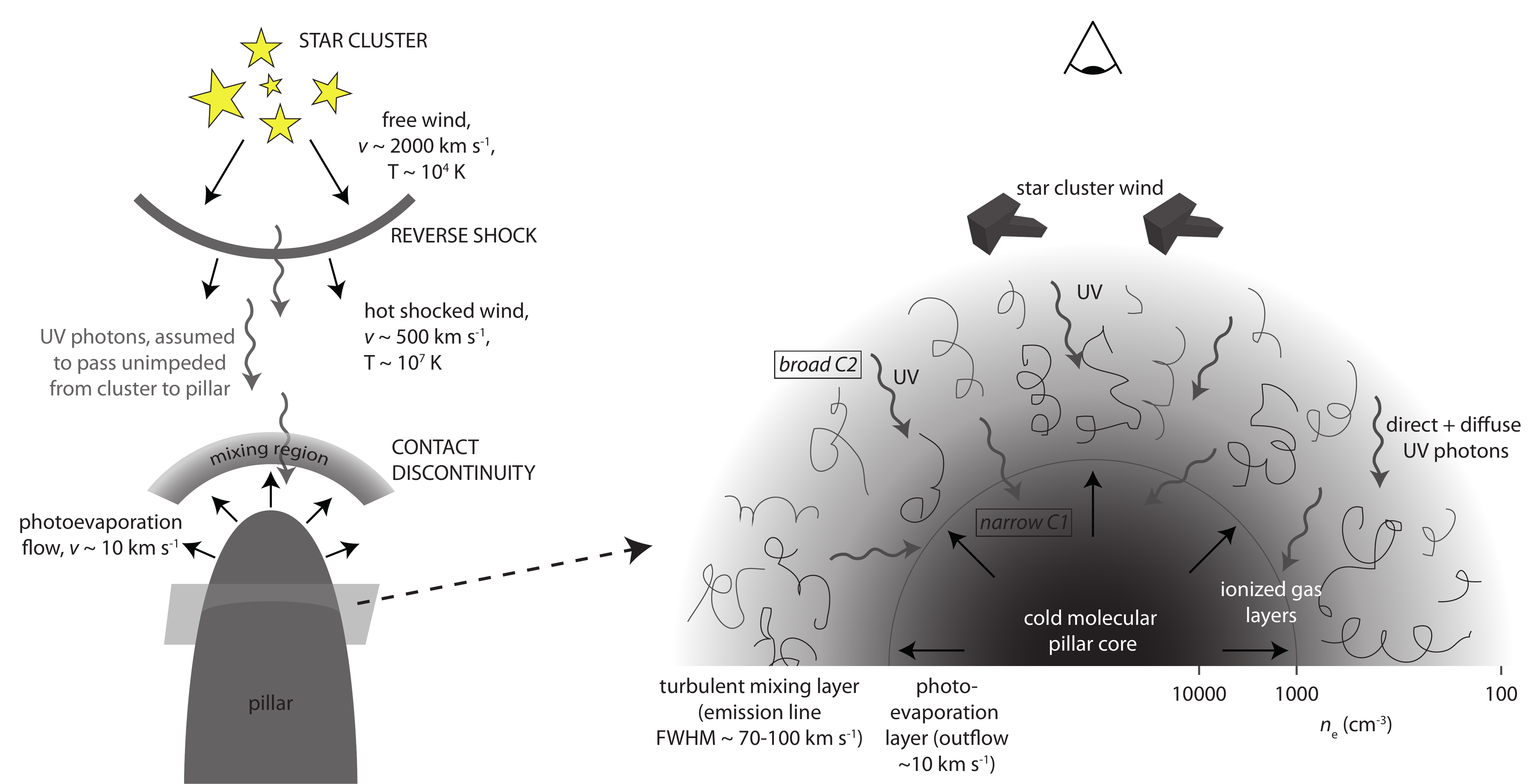}
\caption{Schematic representation of our interpretation of the pillar and its structure. The left panel shows a top-down view of the location of the pillar with respect to the star cluster, showing the cluster wind properties and shock locations, the mixing region and the photoevaporation layer. The right panel shows a cross-section through the pillar showing the different temperature, density, outflow and turbulent mixing layers.}
\label{fig:schematic}
\end{figure*}

\subsection{Schematic model of pillar}
In Fig.~\ref{fig:schematic} we show a schematic representation of the above interpretation of the pillar and its structure. On the left we show a top-down view of the pillar and the star cluster. The cluster emits both ionizing photons and a fast $v\sim 2000$~\kms\ wind \citep[the approximate terminal velocity of a young O star wind, e.g.][]{mokiem07}. The ionizing photons are assumed to travel unimpeded to the pillar surface where they deposit their energy, heating and ionizing the surface gas layers, resulting in a photoevaporative gas outflow of velocity $\sim$10~\kms\ (the sound speed in the $10^4$~K ionized gas). The cluster wind, on the other hand, first shocks against the ambient medium surround the star cluster, and, according to the adiabatic Rankine-Hugoniot relations, decreases in velocity by a factor of 0.25 to $\sim$500~\kms\ and increases in temperature by a factor of $10^3$. 

When the hot, shocked, low density wind then encounters the pillar -- in effect the top of the photoevaporation flow -- a contact discontinuity will be set up. Because the adiabatic sound speed in the hot wind, $(\gamma k_{\rm B}T/\mu)^{1/2}$~\kms\ or $\sim$$370\,(T/10^7~{\rm K})^{1/2}$~\kms, is approximately equal to the speed at which it is travelling, there will be no shock (or only a very weak shock). The two media will come into pressure equilibrium, and a highly turbulent mixing region will form driven by shear, flow (such as Rayleigh-Taylor) and thermal pressure instabilities. It is from this layer that we think the fainter, broad component (C2) is emitted.

The right panel of Fig.~\ref{fig:schematic} shows a more detailed cross-section through the pillar; the cold neutral/molecular core is surrounded by the ionized photoevaporation layer then the turbulent mixing layer. Gas densities decrease from the core outwards, from $>$10\,000~\cmt\ to $<$100~\cmt. The narrow C1 component arises primarily in the denser photoevaporation layer, whereas the broad C2 component arises in the turbulent mixing region. We do not see any evidence in our observations for any material being ablated by the wind and accelerated downstream. We speculate that this is because any ablated material must be heated to temperatures above which optical lines are emitted.

What might the lifetime of such a pillar be? Assuming that the two pillar tips are at a distance of $\sim$1--1.4~pc from the cluster and are of physical size $\sim$10$''$$\sim$0.34~pc, they subtend solid angles of 0.0114--0.0218~str. NGC~3603 produces $\sim$$10^{51}$ ionizing photons per second \citep{kennicutt84, drissen95}, meaning that the tips receive $\sim$$10^{48}$ ionizing photons per second (a lower limit since we are ignoring recombinations between the pillar and the cluster). Taking the equation for the lifetime of an irradiated globule from we can calculate the mass loss rate due to photoevaporation \citep[][their equation 36]{lefloch94}:
\begin{equation}
\dot{M}=10.4\,\left(\frac{\Phi}{10^7~{\rm cm}^{-2}\,{\rm s}^{-1}}\right)^{1/2} \left(\frac{r}{1\,{\rm pc}}\right)^{3/2}~M_\odot~{\rm Myr^{-1}}
\end{equation}
where $\Phi$ is the ionizing flux seen by the pillar in units of $10^7$~cm$^{-2}$~s$^{-1}$, and $r$ is the radius of the pillar tip in units of pc. This gives $\dot{M}\sim300$~\Msol~Myr$^{-1}$. To estimate the mass of the pillars, we can assume a length of 10~pc, radius 0.3~pc and density $10^4$~cm$^{-3}$, giving a total mass of $\sim$700~\Msol; the evaporation time is therefore on the order of 2~Myr. These pillars therefore suffer very rapid dissolution.

\citet{pound98} estimated the lifetime of the pillars in the Eagle Nebula (M16) to be of order 20~Myr, assuming $\dot{M}=2\pi{}R^2\,c\, m_{p}\,n$~\Msol~yr$^{-1}$ (where $R$ is the radius of curvature of the head of the pillar, $c$ is the thermal sound speed, $m_{p}$ is the proton mass, and $n$ is the volume density). However to arrive at this number they seem to have used an unrealistically small value for $R$, and therefore perhaps overestimated their photoevaporation timescales.


\subsection{Shocks and kinematics}
At no point do we measure [N\two]/H$\alpha$ and [S\two]/H$\alpha$ line ratios greater than those that would indicate a strong contribution from shocks \citep[as predicted by shock models][]{allen08}. Since the motions of the gas in the TML are clearly supersonic, shocks must undoubtedly be present, but it is likely that the reason why we do not see their signatures (i.e.\ enhanced forbidden line emission from the post-shock gas) is that in these small spatially resolved regions the shock contribution is completely washed out by contribution from photoionization.

Position-velocity diagrams extracted from the Pos2 pillar and that in NGC~6357 show a consistent offset in radial velocity between the narrow (brighter) components of H$\alpha$ and [N\two] (+ [S\two]) of $\sim$5--8~\kms. We were unable to find a satisfactory explanation for these differences in radial velocities, although the answer most likely lies in the ionization, temperature and density stratification of the pillar surface layers and the location within these that the emission lines originate. 

We do not find any evidence for rotation within the pillars as has been observed in other examples \citep{gahm06} and in simulations \citep{gritschneder10}. Our observations, however, only probe the surface ionized layers of the structures; if internal rotation did exist then the kinematics of the molecular core would need to be probed.

\subsection{The line-of-sight location of the pillars}
In all cases (including the pillar seen in NGC~6357), the pillars are seen in projection against diffusely emitting background gas. The surrounding gas tends to be redshifted compared to the pillar, and this redshifted component is not seen at the location of the pillar leading to the conclusion that, since it is obscured by the pillar, it must be in the background. In Pos1, we find evidence that the diffuse filamentary gas immediately surrounding the pillar (in projection) is also located behind the pillar, but in front of the faint (low density, high excitation) background. These kind of observations begin to allow us to build up a three-dimensional picture of the nebular structure.

\vspace{0.5cm}
The many unanswered questions and issues raised in this study regarding the details of mechanical and radiative interactions with dense clouds or pillars in H\two\ regions highlights the need for new, high-resolution simulations that take into account the effect of the wind, gas flow and turbulence, and ionization effects. We urge the theoretical community to take up this challenge.



\section*{Acknowledgments}
MSW and JED would like to thank Julian Pittard and Steffi Walch for discussions that helped us explain our observations and what might be happening on the pillar surface. The research leading to these results has received funding from the European Community's Seventh Framework Programme (/FP7/2007-2013/) under grant agreement No 229517. Based on observations made with ESO telescopes at the La Silla Paranal Observatory under programme ID 087.C-0106 and 089.C-0016.

\bibliographystyle{mn2e}
\bibliography{/Users/mwestmoq/Dropbox/Work/references}
\bsp

\label{lastpage}
\end{document}